\documentclass{aa}
\usepackage{graphicx}
\usepackage{txfonts}
\usepackage{natbib}
\usepackage{hyperref} 
\usepackage{dsfont}
\usepackage{amsmath}
\usepackage{breqn}
\hypersetup{colorlinks=true,linkcolor=blue,citecolor=blue,urlcolor=blue}
\usepackage{orcidlink}
\usepackage{lscape}
\usepackage{soul}

\begin{document}







\title{ Shaping Galactic Habitability: \\ the impact of stellar migration and gas giants}
\author { E. Spitoni \orcidlink{0000-0001-9715-5727}\inst{1,2}  \thanks {email to: emanuele.spitoni@inaf.it} \and M. Palla \orcidlink{0000-0002-3574-9578}\inst{3,4} \and L. Magrini \orcidlink{0000-0003-4486-6802} \inst{5} \and F. Matteucci \orcidlink{0000-0001-7067-2302}  \inst{1,6,7} \and C. Danielski \orcidlink{0000-0002-3729-2663}  \inst{5} \and M. Tsantaki \orcidlink{0000-0002-0552-2313} \inst{5} \and \\ A. Sozzetti \orcidlink{0000-0002-7504-365X}  \inst{8} \and M. Molero \orcidlink{0000-0002-8854-6547}  \inst{9,1} \and F. Fontani \orcidlink{0000-0003-0348-3418} \inst{5,10,11} \and D. Romano \orcidlink{0000-0002-0845-6171} \inst{4} \and G. Cescutti \orcidlink{0000-0002-3184-9918} \inst{6,1,7} \and L. Silva \orcidlink{0000-0002-7571-5217}\inst{1,2}  }
\institute{INAF - Osservatorio Astronomico di Trieste, via G.B. Tiepolo
 11, 34143, Trieste, Italy  
 \and IFPU, Institute for Fundamental Physics of the Universe, Via Beirut 2, I-34151 Trieste, Italy  
 \and 
  Dipartimento di Fisica e Astronomia “Augusto Righi”, Alma Mater Studiorum, Università di Bologna, Via Gobetti 93/2, I-40129 Bologna, Italy 
  \and   INAF – Osservatorio di Astrofisica e Scienza dello Spazio di Bologna, Via Gobetti 93/3, I-40129 Bologna, Italy 
  \and 
INAF - Osservatorio Astrofisico di Arcetri, Largo E. Fermi 5, 50125, Firenze, Italy
\and Dipartimento di Fisica, Sezione di Astronomia, Università di Trieste, Via G. B. Tiepolo 11, 34143 Trieste, Italy
 \and  INFN Sezione di Trieste, via Valerio 2, 34134 Trieste, Italy
\and INAF – Osservatorio Astrofisico di Torino, Via Osservatorio 20,
10025 Pino Torinese, Italy
\and
Institut f\"ur Kernphysik, Technische Universit\"at Darmstadt, Schlossgartenstr. 2, Darmstadt 64289, Germany
 \and
Max-Planck-Institut f\"{u}r extraterrestrische Physik, Giessenbachstra{\ss}e 1, 85748 Garching bei M\"{u}nchen, Germany
\and
LUX, Observatoire de Paris, PSL Research University, CNRS, Sorbonne
              Universit\'e, F-92190 Meudon, France
}

 \date{Received xxxx / Accepted xxxx}

\abstract {  In exoplanet research, the focus is increasingly on identifying Earth analogs, planets similar in density and habitability potential. As the number of rocky exoplanets grows, parallel discussions have emerged on system architectures and Galactic environments that may support life, drawing comparisons to our own Earth. This has brought renewed attention to the concept of the Galactic Habitable Zone (GHZ) 
as a broader context for interpreting the diversity of planetary environments.
 This study is the first to use detailed chemical evolution models to investigate the impact of stellar migration, modeled through a parametric approach, on the GHZ. Our findings reveal that stellar migration significantly enhances the number of stars capable of hosting habitable planets in the outer Galactic regions, with an increase of up to a factor of five at 18 kpc  relative to a baseline value of unity  at 6 kpc. Furthermore, we explore a novel scenario where the presence of gas giant planets increases the probability for the formation of terrestrial ones. We find that this increased probability is higher in the inner Galactic disc, but is also mitigated by stellar migration. 
In particular, at the present time,  the number of FGK stars hosting terrestrial planets with minimum habitability conditions in the ring centered at 4 kpc is approximately 1.4 times higher than in scenarios where gas giants are assumed to hinder the formation and evolution of Earth-like planets. Without stellar migration, this factor increases to 1.5. 
Even larger ratios are predicted for terrestrial planets orbiting retired A stars, reaching 2.8 in models with stellar migration and 3.3 in models without it.
In conclusion, this study shows that stellar migration predominantly influences the GHZ in the outer Galactic regions, while assuming a positive contribution from gas giants to terrestrial planet formation increases the number of stars capable of hosting habitable planets in the Galactic ring centered at 4 kpc.
}

\titlerunning{Shaping the GHZ}
\authorrunning{Spitoni et al.}
\maketitle

\section{Introduction}

The GHZ was defined by \citet{Gonzalez2001} as the region within the Galaxy where the abundance of metals is sufficiently high to favour the formation and evolution of Earth-like planets. In this context, the chemical evolution of the Galaxy plays a crucial and prominent role, as demonstrated by the work of \citet{lineweaver2004}, which presented, for the first time, a map of the habitability zone as a function of both time and Galactocentric distance. Their study revealed that the most likely regions for the presence of life in our Galaxy are concentrated in an annular region, centred at 8 kpc from the Galactic Centre, 2 kpc wide, which gradually expands over time. They emphasised that, in addition to the presence of a host star and a sufficient quantity of heavy elements to form terrestrial planets, the absence of the disruptive effects of nearby supernova (SN) explosions is a key factor to allow for habitability. Indeed, it is well established that a SN emits intense radiation that can reach the atmosphere of Earth-like planets, causing stratospheric ozone depletion. This allows ultraviolet flux from the host star to penetrate the surface and oceans, potentially damaging genetic material (DNA), which can lead to mutations, cell death, and ultimately, planetary sterilization \citep{Gehrels2003,ellis1995}.
Moreover, effects are also  expected at the global climate level and, consequently, across the entire biosphere—not necessarily limited to the ozone layer \cite[e.g.][]{kirkby2011,svensmark2022,svensmark2023}.
However, the precise effects of SN explosions are not yet fully understood, and the correct method for incorporating them into GHZ models remains an open issue. Even if life on land is destroyed by a nearby SN explosion, it may reappear after several hundred million years. Life shows remarkable resilience, and a cosmic catastrophe may even accelerate the evolution of life forms that are presently unknown \citep[i.e.][]{krug2012,sloan2017}. Moreover, life in deep water could evolve without suffering the destructive effects of a SN explosion.

In recent years, several chemical evolution models have built upon the pioneering work  of \citet{lineweaver2004} presenting new maps of Galactic habitability \citep[e.g.][]{prantzos2008, carigi2013, spitoni2014,spitoni2017,spinelli2021,scherf2024}. In particular, \citet{spitoni2014,spitoni2017}  applied detailed chemical evolution models, including also dust evolution,  to identify habitable zones in the Milky Way for M and FGK stars and Andromeda. 
 Galactic habitability has also been explored in relation to the bulge and the impact of a central AGN \citep[e.g.][]{balbi2020, ambrifi2022}.
 On the other side, cosmological hydrodynamical simulations within the $\Lambda$CDM framework have studied the concept of habitability on larger scales by investigating the types of halos that can produce galactic structures conducive to the formation of habitable planets (\citealt{forgan2017,gobat2016,zackrisson2016,vukotic2016}).

In most of the models discussed above, whether focused on Galactic chemical evolution or based on cosmological frameworks, the possibility of forming planetary systems with terrestrial planets but without gas giants is considered. The presence of gas giants or hot Jupiters could, in principle, disrupt or even destroy Earth-like planets during their formation and evolution. However, there is no clear consensus in the literature regarding the precise effects of gas giants on terrestrial planet formation.
In fact, it is well known that gas giant planets could exert a gravitational influence that promotes the accretion of planetesimals into larger bodies in the inner disk, while simultaneously mitigating excessive dynamical excitation that could otherwise impede their growth. Indeed, \citet{2023he} confirmed that the presence of outer gas giants increases the `gap complexity' in planetary systems, thereby facilitating more efficient formation of terrestrial planets. 
Hence, in this study, we also aim to explore the possibility that giant planets may enhance the probability of terrestrial planet formation.

Moreover, for the first time, this paper considers the effect of stellar migration to develop a habitability map of our Galaxy within the framework of chemical evolution models, using the parametric approach proposed by \citet{frankel2018} and later used in \citet{palla2022}. This is complementary to the work of \cite{baba2024}, who have recently focused on the orbital migration of the solar system. Stellar migration in the Galactic disc has been the subject of several investigations in recent years. \citet{lynden1972} demonstrated that stellar bars and spiral arms cause stars to migrate away from their birth radii through the process known as radial migration. More recently, various explanations of the radial migration mechanism have been proposed, such as bar-induced migration (e.g. \citealt{solway2012,sellwood2014}), and multiple bar/spiral modes (e.g. \citealt{minchev2006,roskar2008}). \citet{minchev2013,minchev2015} and \citet{kubryk2013,kubryk2015} studied the effects of stellar migration on the chemical properties of galaxies, highlighting how the redistribution of stars influences the metallicity gradients and overall chemical evolution.
 Moreover, chemical abundance trends  exhibit significant variations  when considering stars’ guiding or current radii instead of their birth locations \citep{ratcliffe2023, ratcliffe2024}.
In turn, the stellar migration process should also play a crucial role in shaping the conditions necessary for planet formation and the emergence of habitable environments, therefore critically influencing the GHZ. In a related context, \citet{baba2024} emphasized the dynamic nature of Galactic habitability, revealing how a star's journey through the Milky Way profoundly influences its surrounding environment and, consequently, the potential for life. They introduced the concept of "Galactic Habitable Orbits," which considers the evolving structures of the Galaxy and their impact on stellar and planetary systems.

This paper is organised as follows: 
in Section \ref{CEM}, we outline the chemical evolution models used in this study and explain how stellar migration has been incorporated. Section \ref{GHZ_model} presents our GHZ model, while Section \ref{results} details our results. Finally, in Section \ref{conclusions}, we draw our concluding remarks. In Appendices \ref{app:A} and \ref{app:B} more details and results on the reference chemical evolution model and GHZ are reported. In addition, in Appendix \ref{app:data} we present the observational data used for comparison with the metallicity distribution predicted by our models.

\section{The chemical evolution model for the Galactic disc}
\label{CEM}
In Section \ref{molero}, we describe the reference two-infall model adopted in this study.
In Section \ref{ss:migr_model}, we indicate how the stellar migration has been implemented.

\subsection{The reference multi-zone chemical evolution model}
\label{molero}
The adopted chemical evolution model is based on the two-infall model originally developed by \citet{Chiappini1997}. We use the revised version by \citet{molero2023,molero2025}, focusing exclusively on the thick and thin discs, without considering the evolution of the Galactic halo. 
The two-infall model assumes that the Galaxy forms as a result of two distinct gas accretion episodes. The first episode forms the chemical thick disc\footnote{The chemical thick and thin discs are associated to the high-$\alpha$ and low-$\alpha$ sequences observed in the [$\alpha$/Fe] vs. [Fe/H] diagram of Galactic stars, respectively.}, while the second, delayed with respect to the first, leads to the formation of the thin disc. The infalling gas is assumed to be of primordial composition, and the Galactic disc is modeled as independent concentric rings of width \(2\,\mathrm{kpc}\). The star formation rate (SFR) is the one parametrised by the Schmidt-Kennicutt law \citep{kenni1998}:  $\psi(R,t) \propto \nu(R) \,\sigma_{\mathrm{gas}}(R,t)^k$,
where \(\sigma_{\mathrm{gas}}(R,t)\) is the surface gas density, \(k = 1.5\) is the power-law index, and \(\nu\) is the star formation efficiency, expressed in \(\mathrm{Gyr^{-1}}\). The parameter $\nu(R)$ varies with Galactocentric radius according to the prescription by \citet[][see their Table 3]{Palla2020}. The considered initial mass function (IMF) in this work is the one of \citet{Kroupa1993}, i.e., the one specifically derived from star counts in the solar neighbourhood field.
The gas accretion rate in the form of element \(i\), in the two-infall model is:

\begin{equation}
I(R,t) _{i,\mathrm{inf}}= A(R)X_{i,\mathrm{inf}}e^{-t/\tau_1} + \theta(t-t_{\mathrm{max}})B(R)X_{i,\mathrm{inf}}e^{-(t-t_{\mathrm{max}})/\tau_2},
\label{eq_INFALL}
\end{equation}
where  \(X_{i,\mathrm{inf}}\) is the composition of the infalling gas, assumed to be primordial. The timescales \(\tau_1\) and \(\tau_2(R)\) describe the accretion duration for the thick and thin discs, respectively. We adopt \(\tau_1 = 1\,\mathrm{Gyr}\) and allow \(\tau_2(R)\) to vary with radius according to the inside-out formation scenario (e.g., \citealp{Matteucci1989,  Chiappini2001}): $\tau_2(R) = \left(1.033 \,R\,[\mathrm{kpc}] - 1.267\right)\,[\mathrm{Gyr}]$.

We remind the reader that the $\theta$ quantity in  eq. (\ref{eq_INFALL}) is the Heaviside step function. The parameter \(t_{\mathrm{max}}\) denotes the time of maximum accretion onto the thin disc, corresponding to the end of the thick disc formation and the start of the second infall episode. While earlier studies (e.g., \citealp{Chiappini2001, Spitoni2009, Romano2010, Grisoni2018}) typically assumed \(t_{\mathrm{max}} \sim 1\,\mathrm{Gyr}\), recent works suggest a longer delay to reproduce stellar abundances and asteroseismic ages \citep{spitoni2019}. Here, we adopt \(t_{\mathrm{max}} \simeq 3.25\,\mathrm{Gyr}\), following the prescriptions of \citet{Palla2020,palla2022,palla2024} and \citet{molero2023};  
this is  a value also consistent with what was claimed in \citet{nissen2020} and \citet{spitoni2024} studies.
In eq. (\ref{eq_INFALL}), the coefficients $A(R)$ and $B(R)$  are calibrated to reproduce the present-day surface mass densities of the thick and thin discs as functions of radius. Assuming exponential profiles, we adopt:

\begin{equation}
\sigma_{\mathrm{thin}}(R) = \sigma_{8,\mathrm{t}} e^{-(R-R_8) \, [\text{kpc}]/3.5} \text{ and }\sigma_{\mathrm{thick}}(R) = \sigma_{8,\mathrm{T}} e^{-(R-R_8)\, [\text{kpc}]/2.3},
\end{equation}
where $\sigma_{8,\mathrm{t}}$  and $\sigma_{8,\mathrm{T}}$  are the total surface mass densities at 8 kpc. We imposed that $\sigma_{8,\mathrm{t}}+\sigma_{8,\mathrm{T}}=47.1$ M$_{\odot}$ pc$^{-2}$ as suggested by \citet{mckee2015} and assumed by \citet{molero2025}. Following \citet{spitoni2020,spitoni2022},  \citet{mac2017} and \citet{molero2023}, we suppose that the ratio between the two disc components at 8 kpc  is $\sigma_{8,\mathrm{t}}/ \sigma_{8,\mathrm{T}} \sim 4$. 
An important constraint for the chemical evolution model is the present-time stellar surface mass density in the solar vicinity. The proposed  model predicts a value of 34.3 M$_{\odot}$ pc$^{-2}$, in excellent agreement with the value of 33.4 $\pm$ 3 M$_{\odot}$ pc$^{-2}$  suggested by \citet{mckee2015}. Appendix \ref{app:nucleo} presents the adopted nucleosynthesis prescriptions.

\subsection{The implementation of stellar radial migration}
\label{ss:migr_model}

As mentioned in the Introduction, in our analysis we  include 
stellar radial migration in the multi-zone chemical evolution model of the Galactic disc components.
We implement it by including the approach used in \citet{palla2022}. This framework follows the parametric prescriptions proposed by \citet{frankel2018}, where migration is seen as a result of a diffusion process and treated in a parametric way (see also \citealt{Frankel2020}). 
In particular, the probability for a star to be currently at a Galactocentric radius ${\rm R}_f$, given that it was born at $R_i$ and with age $\tau$, can be written as:
    \begin{equation}
    \ln p({R}_f\, |\, R_i, \tau) = \ln(c_3) \, -\frac{({R}_f- {R}_i)^2}{2 \, \sigma_{RM} \, \tau/10\, {\rm Gyr}} ,
    \end{equation}
where $\sigma_{RM}$ is the radial migration strength expressed in kpc and $c_3$ a normalization constant ensuring that stars do not migrate to negative radii (see \citealt{frankel2018} for details). 
For $\sigma_{RM}$ we adopt as reference a value of 3.5 kpc, as used by \citet{palla2022} and very similar to that found in \citet{frankel2018}. However, we also show results for 
$\sigma_{RM}=6$ kpc, in order to test the effect of an extreme case for stellar migration on the GHZ.
As for \citet{palla2022} prescriptions, we assume stellar migration for the two-infall model only after the onset of the second infall ($\sim$ 10 Gyr ago), also similarly to 
\citet{frankel2018}, where migration was not considered for an "old disc" (age > 8 Gyr in their case) component. 

 We recall that \citet{frankel2018,Frankel2020} applied their analysis to fit the observed age–metallicity distribution of low-$\alpha$ red clump stars within Galactocentric radii of 5 to 14 kpc, as derived from  the Apache Point Observatory Galactic Evolution Experiment project (APOGEE DR12, \citealt{alam2015},  DR14 \citealt{abolfathi2018}). In our study, we extend this relation to a broader range, from 3 to 19 kpc. We acknowledge that this methodology may lead to an underestimation of stellar migration from the innermost regions of the Galaxy; however, 
in our parametric approach, this represents the best possible approximation. Moreover, the above-mentioned extreme migration scenario ($\sigma_{RM}=6$) is also meant to overcome such a limitation, showing the effect of a larger number of migrators from the Galaxy inner regions.  In this way, we should also encompass the effects of a $\sigma_{RM}$ varying in time.
    We also assume that radial migration is only a "passive" tracer of chemical evolution, i.e. the metal enrichment of the ISM in the solar vicinity is practically unaffected by the stars born in other regions.

The choice of the above-mentioned prescriptions to model the radial migration effects is driven by the analogue chemical evolution model framework adopted in \citet{palla2022}, which is already tested in the light of abundance data in the solar vicinity. In particular, \citet{palla2022} used a chemical evolution model characterised by a two-infall scenario with a significant delay (3.25 Gyr) between the accretion episodes, reproducing the abundance diagrams as traced by the AMBRE:HARPS observational data sample \citep{santosperal20,santosperal2021} within the AMBRE project \citep{delaverny2013}, assuming a radial migration strength of $\sigma_{RM}=3.5$.
This assures us to probe the effects of stellar migration on the GHZ in a reasonable and solid physical framework, already tested in the context of Galactic observations.

\section{The Galactic habitable zone model}
\label{GHZ_model}

The main goal of this work is to provide updated GHZ maps compared to those presented in \citet{spitoni2014,spitoni2017} by taking into account the effects of stellar migration and different prescriptions for gas giants.
In the following, we present  the main ingredients required to obtain a map of Galactic habitability as a function of the Galactic time and Galactocentric radius using  the formalism introduced by  \citet{spitoni2014}.

\subsection{The GHZ model without stellar migration}
\label{GHZ_nomigr}
We define $P_E$ as the probability of forming a terrestrial planet around a star. 
\citet{buchhave2012}, analysing data from the Kepler mission, found that the occurrence rate of Earth-sized planets around single stars remains nearly independent of the host star’s metallicity for [M/H] values below 0.5 dex (see also \citealt{buchhave2018}). This finding has  been confirmed by other analyses based on Kepler and K2 \citep[e.g.][]{petigura2018, zink2023}. A minimum metallicity threshold for planet formation was identified by \citet{johnson_li2012} by comparing the timescales of dust grain growth and settling, with that of disk photoevaporation. Their theoretical study, which considered a wide range of circumstellar disc models and dust grain properties, indicated the existence of a critical metallicity  above which planets can form as a function of the distance  at which the planet orbits its host star. In particular, their calculations support the hypothesis that the first Earth-like planets likely formed from circumstellar discs with  [Fe/H] $\simeq$ -1.0 dex, but planets are unlikely to form at lower metallicities. 
  In order to be consistent with these findings, \citet{spitoni2017} assumed a probability of 0.4 for the formation of Earth-like planets around stars with [Fe/H] > -1 dex, while setting $P_E = 0$ for stars with lower metallicity. The specific choice of $P_E=0.4$ aligns with the metallicity-integrated probability proposed by \citet{lineweaver2004}.

In \citet{spitoni2014,spitoni2017}, the authors explored the scenario in which gaseous giant planets orbiting the same host star can disrupt terrestrial planets, particularly during their migration around the host star. \citet{armitage2003} highlighted the potential dangers posed by gas giant migration to Earth-like planets' formation, suggesting that such planets are more likely to form in systems where massive giants underwent minimal migration. Similarly, \citet{matsumura2013} investigated the orbital evolution of terrestrial planets in dynamically unstable systems with gas giants, showing that even Earth-like planets located far from the giants can be ejected.

We adopt the probability of formation of a gaseous giant planet as the function of the host star mass and  iron abundance from \citet{ghezzi2018}:
\begin{equation}
 P_{GGP}\left(M_{\star}, \mbox{[Fe/H}]\right)= 0.085^{+0.008}_{-0.010} \left( \frac{M_{\star}}{M_{\odot}} \right)^{1.05^{+0.28}_{-0.24}} 10^{1.05^{+0.21}_{-0.17}\mbox{[Fe/H]}}.
\label{ghezzi}
\end{equation}
In particular, this relation provides a good fit for the occurrence frequency of gas giant planets across a wide range of stellar host masses, including M, FGK, and retired A  stars. To take into account of  the quantity $P_{GGP}$ in our GHZ model, we weighted   the term $(M_{\star}/M_{\odot})^{1.05}$  on the \citet{Kroupa1993}  IMF for the different mass ranges spanned by the considered stellar types \footnote{We note that the lower limit for M stars is set at 0.1 $M_{\odot}$, consistent with the minimum mass adopted in the IMF used in our chemical evolution model.}.
Assuming  that the range of masses spanned by $M$-type stars is  
$0.1 \leq (M_{\star}/M_{\odot}) \leq 0.45$, for FGK stars  is $0.45 \leq (M_{\star}/M_{\odot}) \leq 1.40$ and for  retired A stars  is $1.40 \leq (M_{\star}/M_{\odot})  \leq 2.40$, we have that  the IMF weighted  $P_{GGP}$ becomes:   
\begin{equation}
\tiny
<P_{GGP}\left(M_{\star}, \mbox{[Fe/H]}\right)>_{IMF} = 
\begin{cases}
M, &  0.085 \times 0.205   \times   10^{1.05 {\mbox{[Fe/H]}}}\\
FGK, &  0.085 \times 0.723   \times   10^{1.05 {\mbox{[Fe/H]}}}\\
retired \, A, &  0.085 \times 1.834   \times   10^{1.05 {\mbox{[Fe/H]}}}.
\end{cases}
\end{equation}
\begin{figure}
	      \centering
       \includegraphics[scale=0.35]{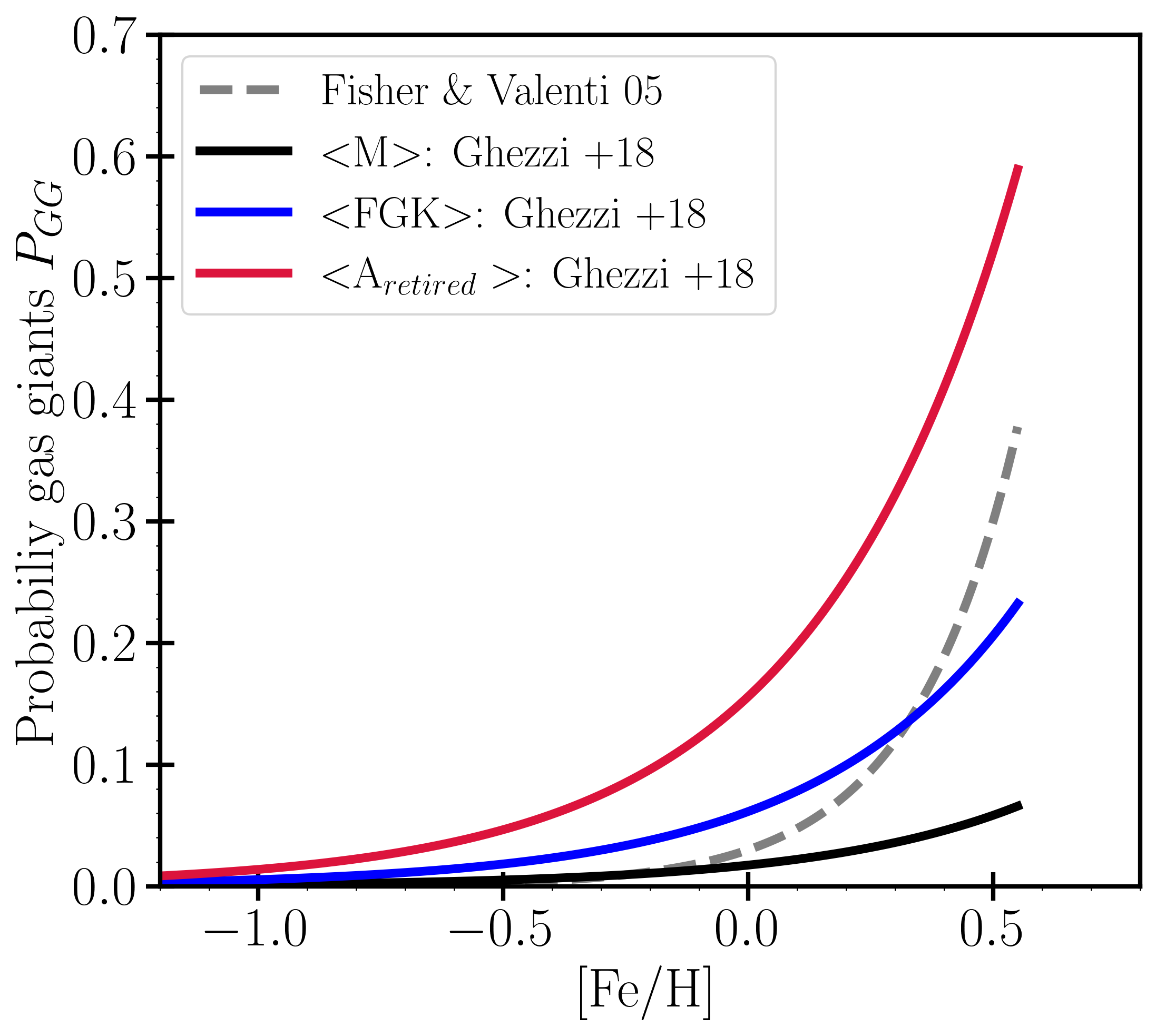}
\includegraphics[scale=0.35]{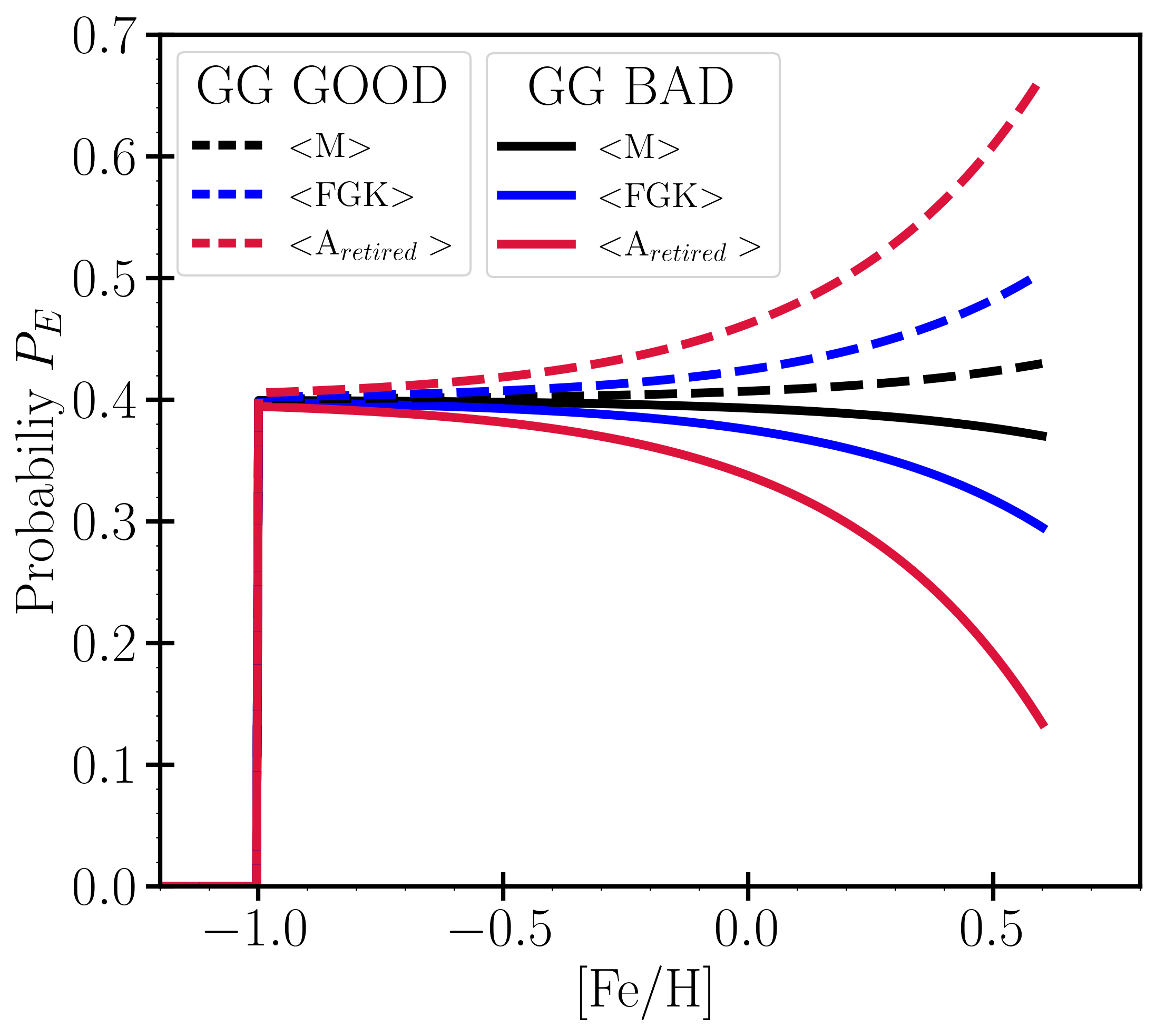}
    \caption{  The probability of forming  planets as a function of  metallicity and stellar type. {\it Upper Panel}:  probabilities of forming  gas giant planets as a function of [Fe/H] for M, FGK and retired A  proposed by \citet{ghezzi2018} are reported with the black, blue and red lines, respectively. The probabilities have been IMF weighted  as described in Section \ref{GHZ_nomigr}. \cite{fischer2005} relation is also shown with the dashed grey line.
 {\it Lower Panel}:    solid lines represent the probability $P_E$ of forming Earth-like planets and not gas giants  (the "GG BAD" case) as a function of [Fe/H], and indicated in eq. (\ref{PE_bad}) for different stellar spectral types. In contrast, the dashed lines illustrate the new scenario discussed in Section  \ref{sect_model_good} and called the "GG GOOD" case, which assumes that the presence of gas giants enhances the likelihood of terrestrial planet formation. All the cases are computed using the gas giant probabilities of \citet{ghezzi2018}.  }
    \label{GOOD}
\end{figure} 
\begin{figure*}
	      \centering
          \sidecaption
\includegraphics[scale=0.35]{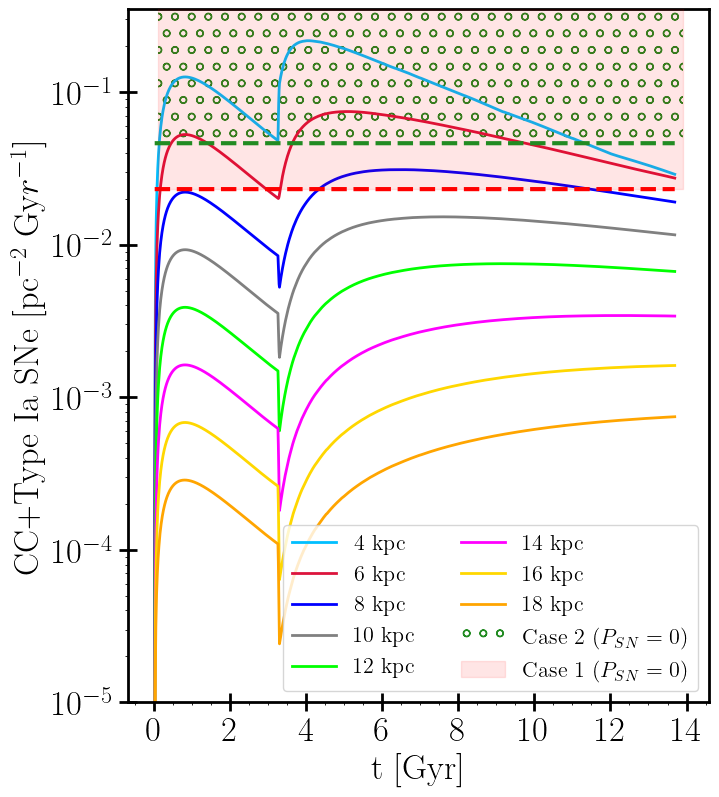}
      \includegraphics[scale=0.35]{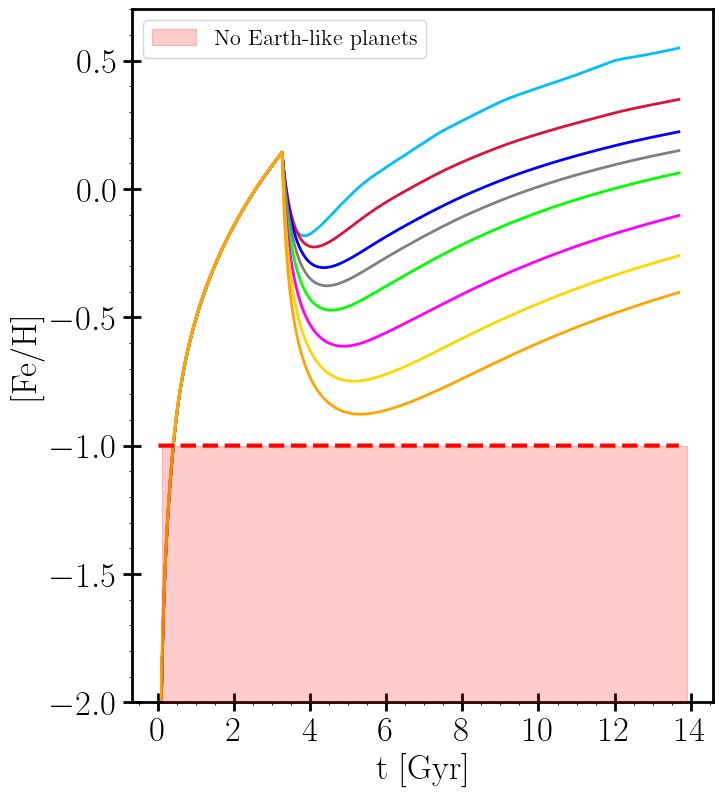}
  \caption{ Predictions of  our multi-zone chemical evolution model presented in Section 
  \ref{molero} as a function of the evolutionary time $t$ and the Galactocentric distance.  {\it Left Panel}:  evolution of the total (CC+Type Ia)  SN rates. The red and green dashed horizontal lines indicate <$R_{SN,\,\odot}$>  (case 1) and 2$\times$<$R_{SN,\,\odot}$> (case 2) values, respectively, representing the minimum SN rate thresholds tested for SN explosion-induced destruction effects in this study.
{\it Right Panel}: Predicted age-metallicity relation. The red-shaded region marks [Fe/H] < -1 dex where, as suggested by \citet{johnson_li2012}, conditions are considered unfavourable for the formation of terrestrial planets.}
		\label{SFR_ref}
\end{figure*} 
In the upper panel of Fig. \ref{GOOD}, we show the  <$P_{GGP}\left(M_{\star}, \mbox{[Fe/H]}\right)$>$_{IMF}$ quantity for M, FGK and retired A stars. As a reference, in the same plot, we draw the \citet{fischer2005} $P_{GG}$ probability where no dependence on the stellar mass was considered.
Hence,  the probability of forming an Earth-like planet, $P_{E}$, as a function of the host star [Fe/H] abundance, assuming no formation of gas giants, is:
\begin{equation}
 P_E\left(\mbox{[Fe/H}]\right)= 0.4 \times\left(1-<P_{GGP}\left(M_{\star}, \mbox{[Fe/H}]\right)>_{IMF}\right),
\label{PE_bad}
\end{equation}
as shown in  the lower panel of Fig. \ref{GOOD} by the solid lines labelled as "GG BAD" for M, FGK and retired A stars. It is worth commenting on the baseline frequency of Earth-like planets (excluding giant planets). We acknowledge that several studies in the literature have focused on deriving this value in different cases. For instance, \citet{bergsten2022} estimated a lower occurrence rate for Sun-like stars, around 0.15–0.2, while for M dwarfs, this value appears to be higher, in the range of 0.5–0.8. However, to ensure consistency with previous works on the modeling of the GHZ \citep{lineweaver2004, prantzos2008, spitoni2014, spitoni2017}, we adopt a uniform value of 0.4 for all stellar types. Having expressed the probability $P_{E}$, we can then define the total number of stars formed at a certain time $t$ and Galactocentric distance $R$  hosting Earth-like planets with minimum Habitability Conditions $N_{\star \, mHC}(R,t)$, as:
\begin{equation}
N_{\star \, mHC}(R,t)=P_{GHZ}(R,t) \times N_{\star tot}(R,t), 
\label{Ns}
\end{equation} 
where  $N_{\star tot}(R,t)$ is the total number of stars formed  up to  time $t$ (and still alive at that evolutionary time) at the Galactocentric distance $R$, while
$P_{GHZ}(R,t)$  is  the fraction of all stars
having Earths (but no gas giants) which survived SN explosions as a function of the Galactocentric distance and time. 
This latter quantity can be expressed as:
\begin{equation}
P_{GHZ}(R,t)= \frac{\int_0^t \psi (R,t') P_E (R,t') P_{SN}(R,t')\ dt'}{\int_0^t \psi(R,t')\ dt'},
\label{GHZ}
\end{equation}
and must be interpreted  as the relative
probability of having minimum habitability conditions around one star at a given position (see \citealp{prantzos2008}).
In eq. (\ref{GHZ}), $\psi(R,t')$ is the star formation rate (SFR) at the time $t'$ and Galactocentric distance $R$; $P_{SN}(R,t')$ is the probability of surviving a SN explosion  (other catastrophic events, such as nearby nova outbursts or compact object mergers, are not taken into account because of the rarity of their occurrence). 
Due to uncertainties on the actual impact of SNe on life destruction, we consider two possible scenarios, called Case 1 and Case 2, as presented in \citet{spitoni2014}. In these scenarios, SN destruction is assumed to be effective if the total SN rate (i.e. CC + Type Ia) at any given time and location exceeds the average SN rate in the solar vicinity over the past 4.5 Gyr of the Milky Way's life (<$R_{SN,\,\odot}$>) for Case 1, and if it is larger than 2$\times$<$R_{SN,\,\odot}$> for Case 2.
Therefore, the associated probability $P_{SN}(R,t)$ is:
\begin{equation}
P_{SN}(R,t) = 
\begin{cases}
0, & \text{if } \ \ \ \ R_{SN}(R,t)\ >\  <R_{SN, \,\odot}>\ \ \ \  \ \ \text{ (Case 1)} \\
  & \lor \,\ \ \  \ R_{SN}(R,t)\ >  2 \times <R_{SN,\,\odot}> \, \text{ (Case  2)} \\
1, & \text{otherwise}
\end{cases}
\end{equation}
For <$R_{SN,\,\odot}$>, we adopt the value of 0.023 Gyr$^{-1}$ pc$^{-2}$ as the results of the reference chemical evolution model presented in Section \ref{molero}.
To understand how the average <$R_{SN,\,\odot}$> compares to the SN rates at different Galactocentric radii, the  
left panel of Fig. \ref{SFR_ref} presents the Case 1 and Case 2 thresholds alongside the predicted time evolution of total SN rates.
It is important to stress that in this work we are exploring the  possibility of Earth-like planets located in regions safe from SN damage. However, the number of life-hosting planets among them may be significantly reduced when considering additional key requirements for life development, specifically N$_2$-O$_2$-dominated atmospheres with constrained CO$_2$ levels, as recently discussed in \citet{scherf2024}.
In addition, it is worth noting that a detailed analysis of the effects of SN explosions on the GHZ is beyond the scope of this paper. Our primary aim is to highlight the impact of stellar migration on previously presented models. A more precise and comprehensive treatment of SN-induced disruptions to habitability will be addressed in future work.

\begin{table}
\begin{center}
\tiny
\caption{Summary of the models presented in this study predicting the number of stars hosting terrestrial planets with life ($N_{\star, mHC}$)}.
\label{tab_A}
\begin{tabular}{|c|cccc|}
\hline
  \hline
 & & & &  \\
Models  & $P_{E}$([Fe/H])  & SN effects & Stellar& $\sigma_{RM}$\\
 &   &  & migration& \\
  & see Fig. \ref{GOOD}  &  & & [kpc] \\
 
\hline
 & & & &  \\
Model 1  &  GG BAD  & case 2 &  No &/ \\
 & & & &  \\
 Model 2  &  GG BAD  & case 1 &  No &/ \\
 & & & &  \\
 Model 3  &  GG BAD  & case 2 &  Yes & 3.5 \\
 & & & &  \\
 Model 4  &  GG BAD  & case 2 &  Yes & 6.0 \\
 & & & &  \\
 Model 5  &  GG GOOD  & case 2 &  Yes & 3.5 \\
 & & & &  \\
 Model 6  &  GG GOOD  & case 2 &  Yes & 6.0 \\
 & & & &  \\
 Model 7  &  GG GOOD  & case 2 &  No & / \\
 & & & &  \\
\hline
\end{tabular}
\end{center}
\tablefoot{Model names are reported in the first column. The adopted probability $P_E$ of finding terrestrial planets  as the function of the metallicity in the presence of gas giants is indicated in the second column.    The SN damage model (case 1 or case 2) is reported in the third column. The presence of the stellar migration and the value of  the migration strength $\sigma_{RM}$ are indicated in the fourth and last columns, respectively.}
\end{table}

\subsection{GHZ formalism in the presence of the stellar migration}

In the presence of the stellar migration, the definition of  $P_{GHZ}$ and the formalism to compute $N_{\star, mHC}$ must be revisited to account for the increased complexity added to the system.
Here, we adopt 
the following expression: 
\begin{equation}
N_{\star \, mHC}(R_f,t)=\sum_{R_{i}}\left(P_{GHZ}(R_{i} \rightarrow R_f,t)  \times N_{\star \,}(R_{i} \rightarrow R_f,t) \right),
\label{Ns_m}
\end{equation}
where $P_{GHZ}(R_{i} \rightarrow R_f,t)$ is defined as:
\begin{equation}
P_{GHZ}(R_{i} \rightarrow R_f,t)={\frac{\int_0^t \psi (R_i,t') P_E (R_i,t') P_{SN}(R_f,t') \ dt'}{\int_0^t \psi(R_i,t') \ dt'} }.
\label{GHZ_m}
\end{equation}
The quantity $P_{GHZ}(R_{i} \rightarrow R_f,t)$ expresses the fraction of all stars born at the Galactocentric distance $R_i$ that have migrated to $R_f$ by the time $t$ and could potentially host a terrestrial planet. 
In eq. \eqref{GHZ_m}, the abundance ratio [Fe/H] used to compute the probability $P_E$ (see eq. \ref{PE_bad}), is determined by the value predicted by the chemical evolution model at the star birth radius $R_i$. 
This is because the stellar metallicity is inherited by the ISM abundance at stellar birthplace and birthtime \citep[e.g.][]{trimble1996,pagel2006,matteucci2021}. 
 However, the SN explosion damage effect, represented by $P_{SN}$, must be evaluated at the star final location $R_f$, as SN explosions happening at time $t$ are impacting planetary systems in their surroundings, without relations with stellar planetary host system birthplace. We emphasize that for stars formed in situ ($R_i = R_f$), the expressions given in eqs. (\ref{Ns_m}) and (\ref{GHZ_m}) are equivalent to those in eqs. (\ref{Ns}) and (\ref{GHZ}) presented in Section \ref{GHZ_nomigr} for the reference model without stellar migration.
In summary, the total number of stars hosting habitable terrestrial planets that have migrated from $R_i$  to  $R_f$ by the evolutionary time $t$, is obtained by multiplying  $P_{GHZ}(R_{i} \rightarrow R_f,t)$  by the number of migrating stars, $N_{\star}(R_{i} \rightarrow R_f,t)$. In order to account for all stars that have 
reached $R_f$  with the potential to support life, we sum over the different birth radii  $R_i$, as expressed in eq. (\ref{Ns_m}).
 
\begin{figure}
	      \centering
\includegraphics[scale=0.29]{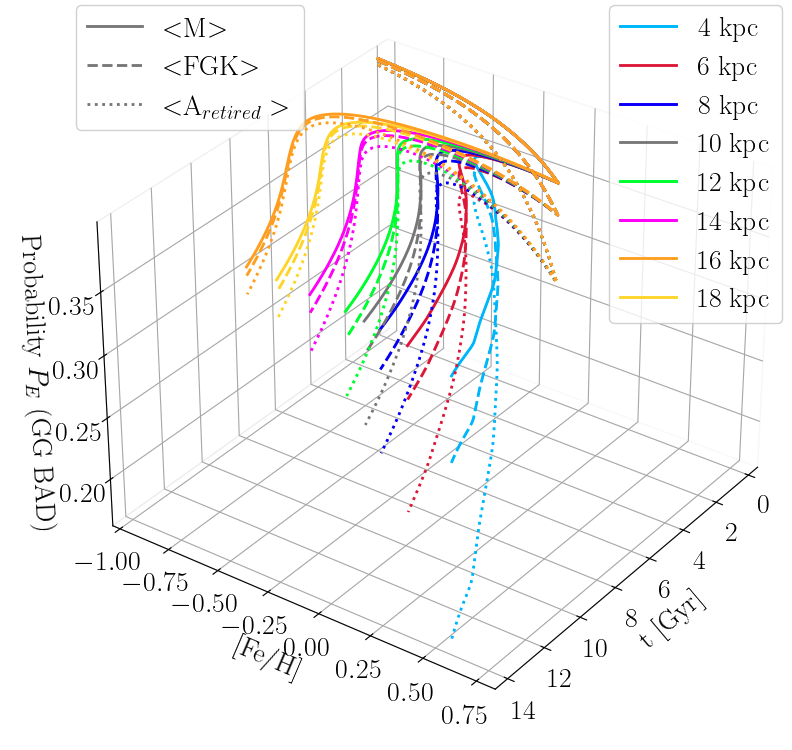}
    \caption{ Evolution of the  probability $P_E$ of forming Earth-like planets and not gas giants (the   "GG BAD" case  reported in eq. \ref{PE_bad} and  in Fig. \ref{GOOD})    as predicted by our reference multi-zone chemical evolution model presented in Section \ref{molero} in the  3D space formed by the $P_E$, [Fe/H] and evolutionary time $t$ for different stellar types.  }
    \label{multi_PE}
\end{figure} 

\begin{figure}
	      \centering
\includegraphics[scale=0.42]{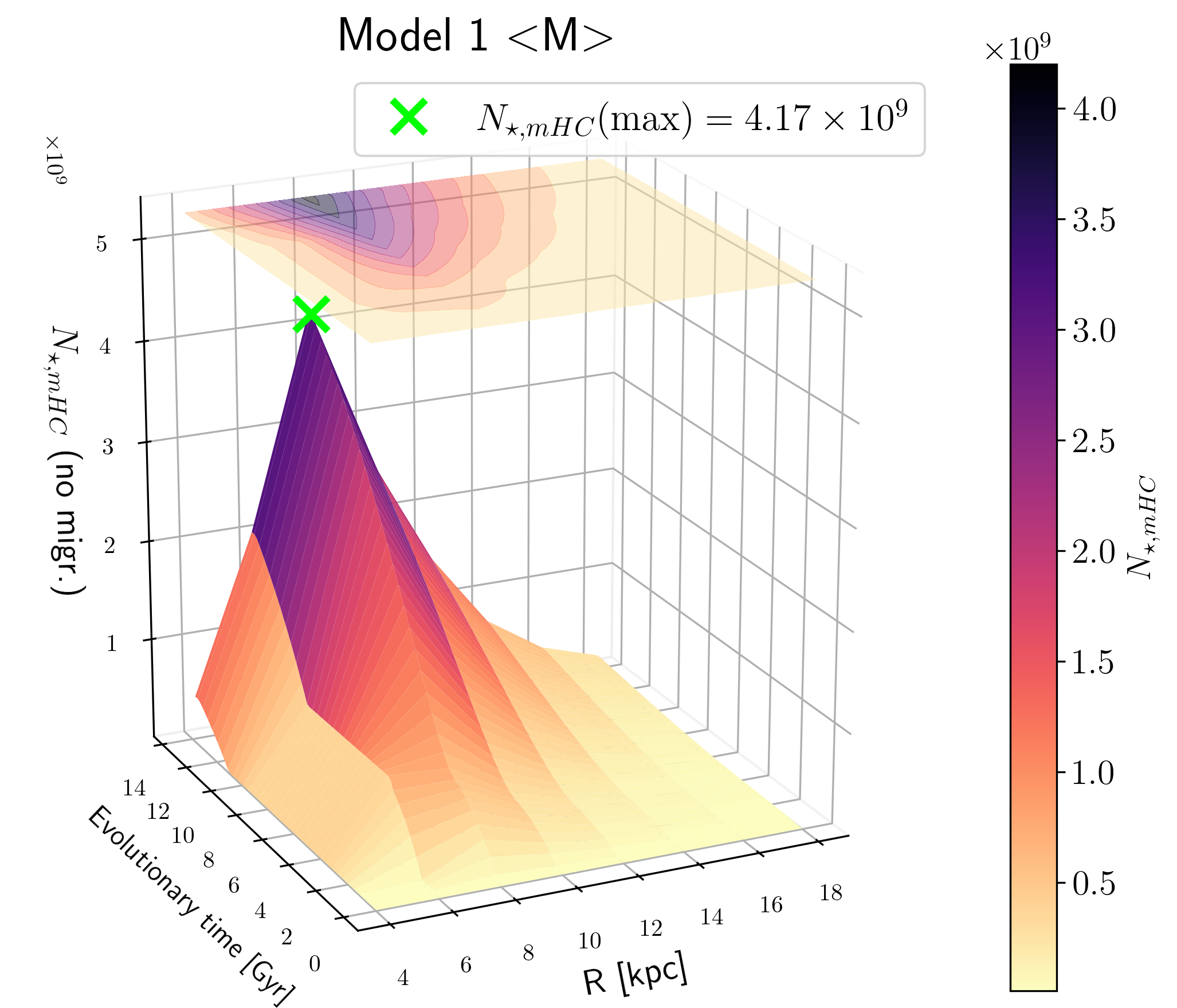}
  \includegraphics[scale=0.42]{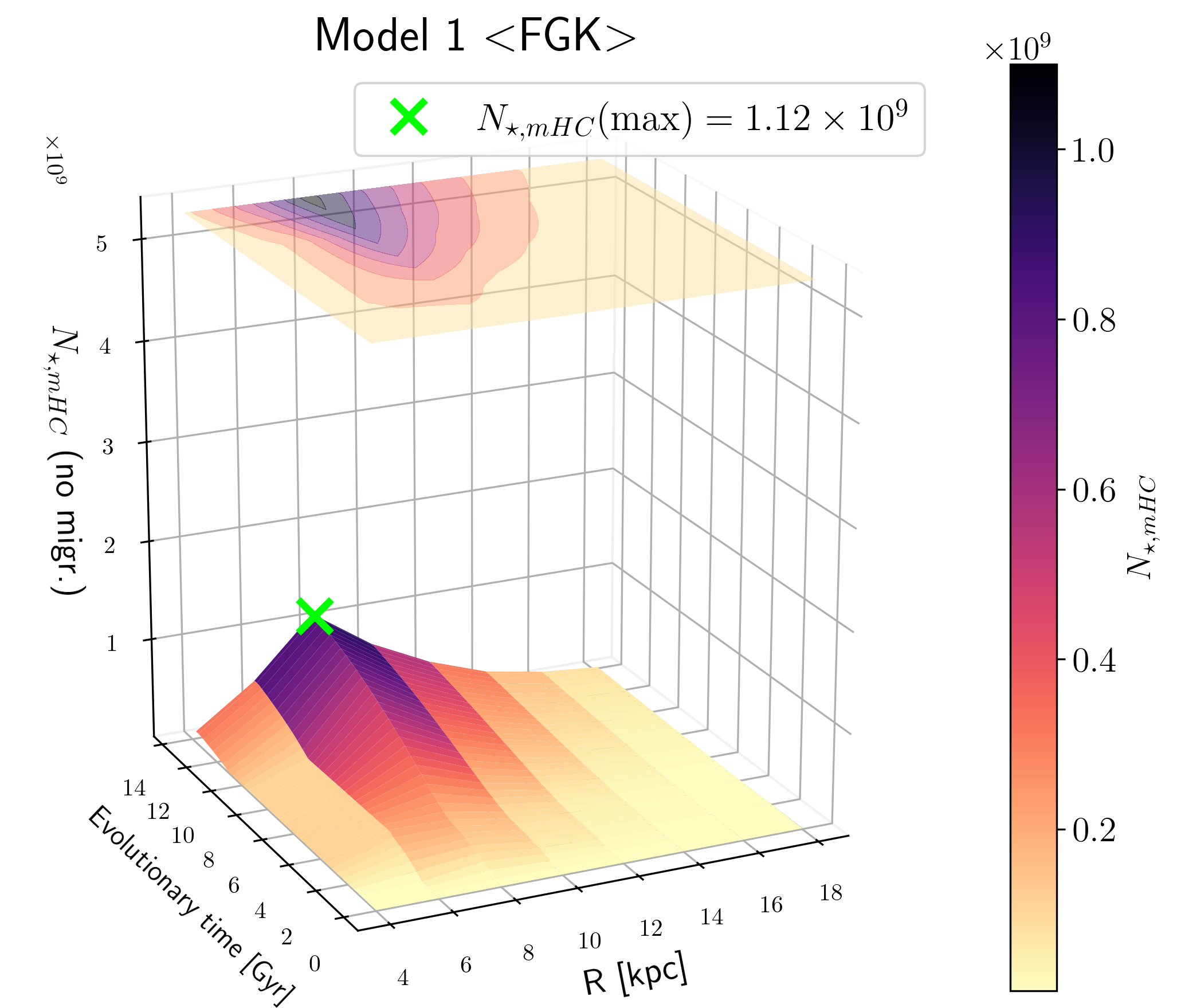} 
  \includegraphics[scale=0.42]{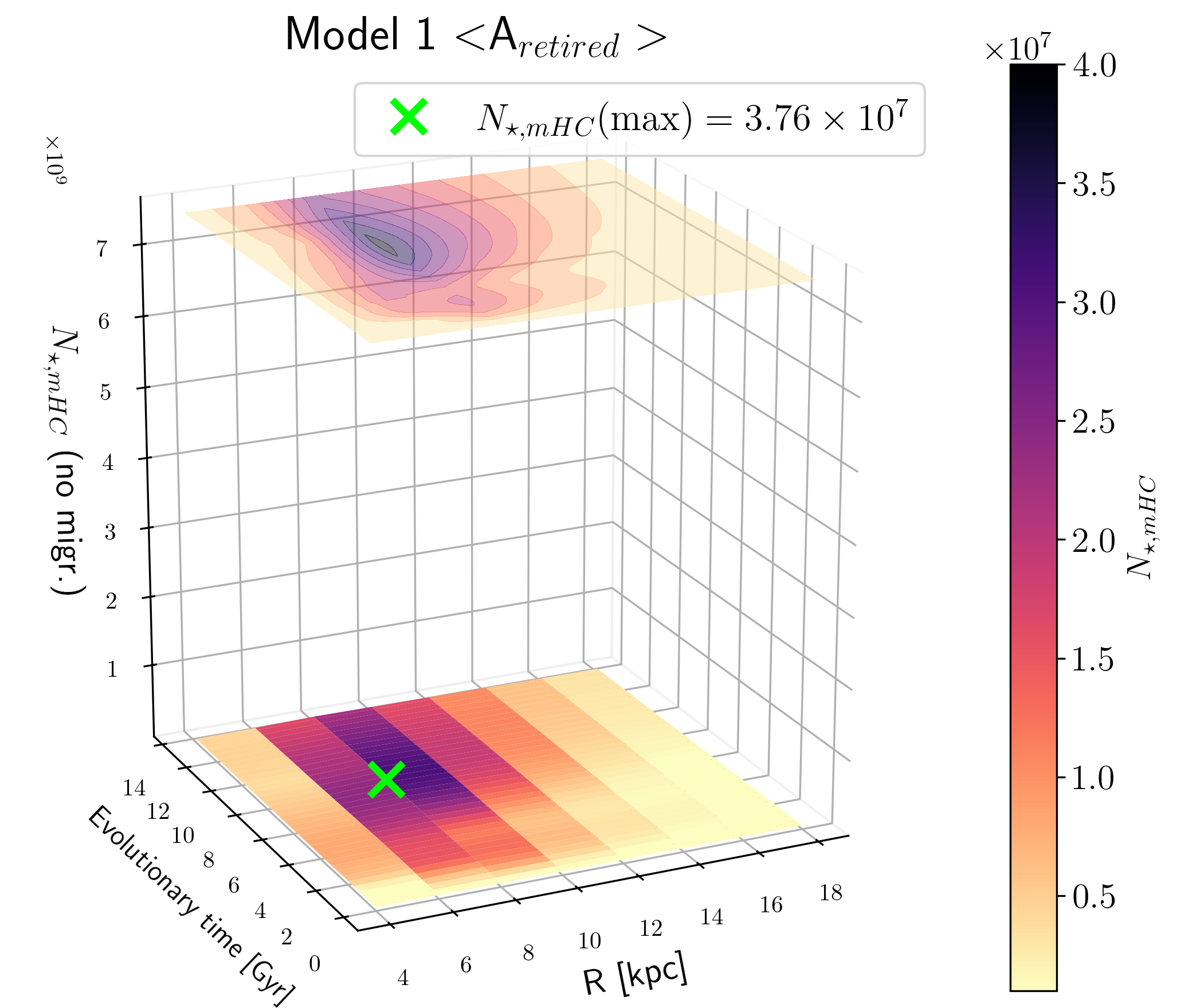}
     \caption{  Total number of different spectral stellar types  hosting Earth-like planets with minimum habitability conditions ($N_{\star, mHC}$ in eq. \ref{GHZ}) as a function of Galactic distance and time predicted by the chemical evolution model without stellar migration (Model 1, see Section \ref{GHZ_nomigr} and Table \ref{tab_A}). The values of ($N_{\star, mHC}$) are computed within concentric rings of 2 kpc width. Results for M stars are drawn in the upper panel.  FGK and retired A are shown  in the middle  and the lower ones, respectively. 
     }
		\label{models_MFGKA}
\end{figure}

\begin{figure}
	      \centering
     \includegraphics[scale=0.48]{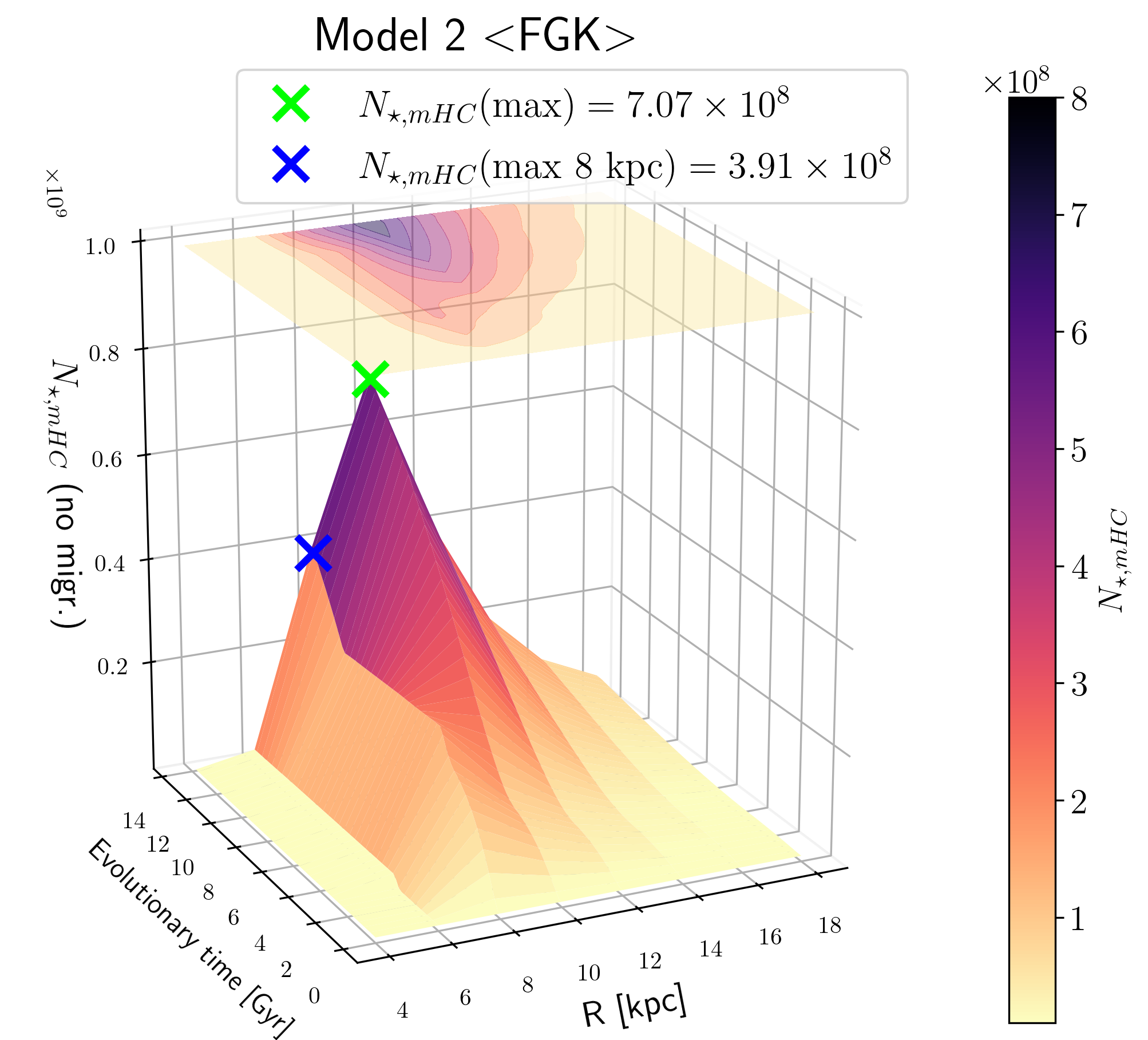}
     \caption{Total number of FGK stars   with minimum habitability conditions ($N_{\star, mHC}$ in eq. \ref{GHZ}) as a function of Galactocentric distance and Galactic time predicted by the Model 2, i.e. Case 1 SN damage scenario  and no stellar migration (see Section \ref{GHZ_nomigr} and Table \ref{tab_A}). }
		\label{models1.2}
\end{figure} 

\subsection{Gas Giants  as catalysts for Terrestrial Planet Formation}
\label{sect_model_good}

In previous GHZ models gas giants have been considered an important hazard for  the creation and evolution of terrestrial planets. 
For instance, \citet{walsh2011} discussed how Jupiter’s migration sculpted the inner solar system, limiting material available for terrestrial planets.  
On the other hand, gas giants clear material in their orbital vicinity, creating gaps in the protoplanetary disk. This can enhance planet formation by concentrating planetesimals and embryos into specific regions, as described in the “Nice Model” and similar frameworks \citep[e.g.][]{tsiganis2005}. For instance, simulations show this process can help stabilize inner regions for terrestrial planet growth \citep{morbidelli2012}.
The gravitational influence of gas giants promotes the accretion of planetesimals into larger bodies in the inner disk while also mitigating excessive dynamical excitation that might hinder their formation. \citet{Zhu2018} found a positive correlation between the presence of an Earth-like planet and the existence of a cold giant within the same system. \citet{2023he} confirmed that the presence of outer giants increases the “gap complexity”, favouring more efficient terrestrial planet formation.
Gas giants can also scatter icy bodies from the outer disk inward, delivering essential volatiles like water to form terrestrial planets. This is supported by isotopic analysis of Earth’s water, which aligns with an outer Solar System origin, corroborated by \citet{obrien2018}.

All these findings highlight the dual role of gas giants as either facilitators or potential disruptors of Earth-like planet formation, depending on their location, mass, and dynamical evolution. 
For this reason, in this study, we also explore the opposite scenario to the one proposed in \citet{spitoni2014,spitoni2017}, namely by considering the probability of forming an Earth-like planet, $P_E$, in the extreme case where the formation is pumped up by the presence of gas giant planets.
The dashed lines in Fig. \ref{GOOD} 
represent this extreme case for different stellar spectral types, which  can be expressed as follows:
\begin{equation}
 P_E\left(\mbox{[Fe/H}]\right)= 0.4 \times\left(1+<P_{GGP}\left(M_{\star}, \mbox{[Fe/H}]\right)>_{IMF}\right).
\label{PE_good}
\end{equation}
Throughout this article, we refer to this case as “GG GOOD".
In Fig. \ref{GOOD}, we note that differences between "GG BAD" and "GG GOOD" scenarios are more pronounced for retired A stars, particularly at super-solar values.
For the first time in GHZ map modelling, we will test this assumption using chemical evolution models, showing the results in Section \ref{GG_GOOD_sect}.
In Table \ref{tab_A}, we summarise all the  GHZ models we will consider in this study obtained by varying:
the probability of finding terrestrial planets as a function of the metallicity in the presence of gas giants (GG GOOD or GG BAD), the SN damage scenario (case 1 or case 2), and the stellar migration parameters.

\section{Results}
\label{results}

In this Section, we report our main results
based on the model prescriptions introduced in Sections \ref{CEM} and \ref{GHZ_model}. We present and discuss the results for the ``classical'' chemical evolution model, which does not account for stellar migration, in Section \ref{sec_results_nomigr}. Then, in Section \ref{res_migr}, we report the main effects of stellar migration on the computation of the GHZ. Finally, GHZ maps obtained under the assumption that gas giant planets favour the formation of Earth-like planets are shown in Section \ref{GG_GOOD_sect}.

\subsection{Reference model results without stellar migration}
\label{sec_results_nomigr}

In Fig. \ref{SFR_ref}, we show the temporal evolution of total SN rates and the age-metallicity relation of the multizone model presented in Section \ref{molero}. Concerning the SN rates, we also indicate the <$R_{SN,\,\odot}$>  and  2$\times$<$R_{SN,\,\odot}$> quantities defined in Section \ref{GHZ_model}. 
For most of the thin disc evolution in the solar vicinity (the annular region centered at 8~kpc), the calculated SN rates exceed the average threshold,  <$R_{SN,\,\odot}$>. However, when the higher limit of 2$\times$<$R_{SN,\,\odot}$> is imposed, the evolution remains entirely below this threshold.
\begin{figure}
	      \centering
          \sidecaption
     \includegraphics[scale=0.35]{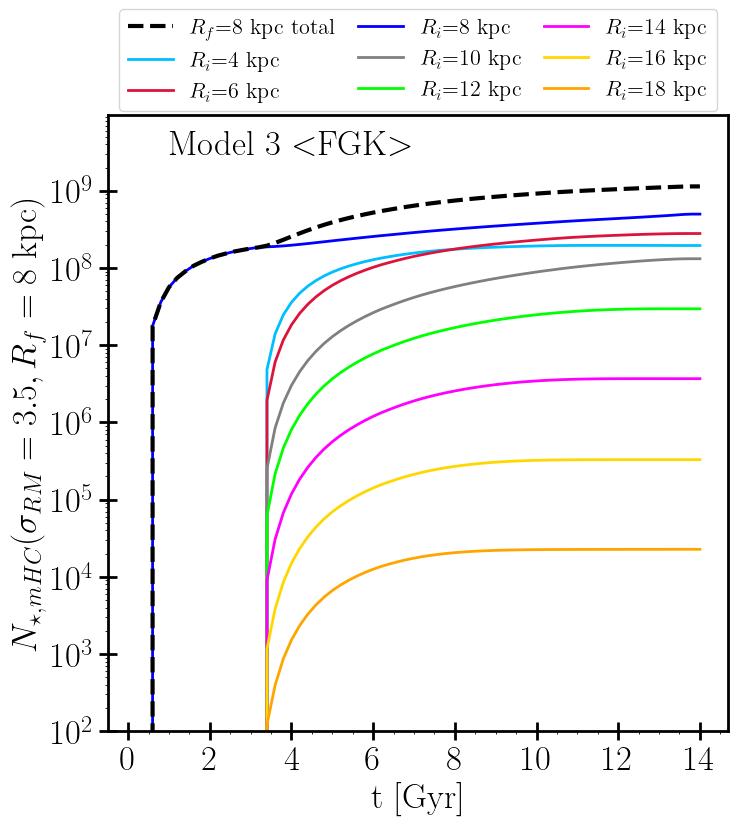}
      \includegraphics[scale=0.35]{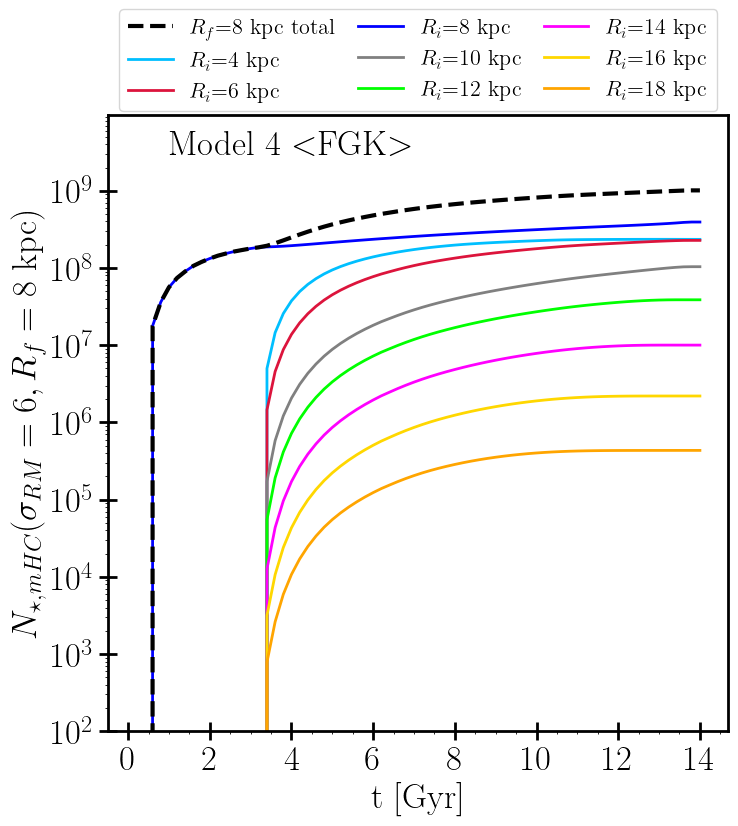}
  \caption{ Temporal evolution of the total number of  FGK stars hosting minimum conditions for life ($N_{\star, mHC}$) in the solar vicinity (black dashed lines). Both panels show the contribution from stars born in situ within the solar annulus ($R_i = R_f = 8$ kpc, dark blue lines) as well as those migrating from different Galactic regions ($R_i \neq R_f$). We consider the Case 2 SN destruction scenario in the presence of stellar migration, with radial migration strength of $\sigma_{RM} = 3.5$ kpc (Model 3, upper panel) and $\sigma_{RM} = 6$ kpc (Model 4, lower panel), respectively. }
		\label{rirf8}
\end{figure} 
 \begin{figure*}
	   \sidecaption
            \includegraphics[scale=0.34]{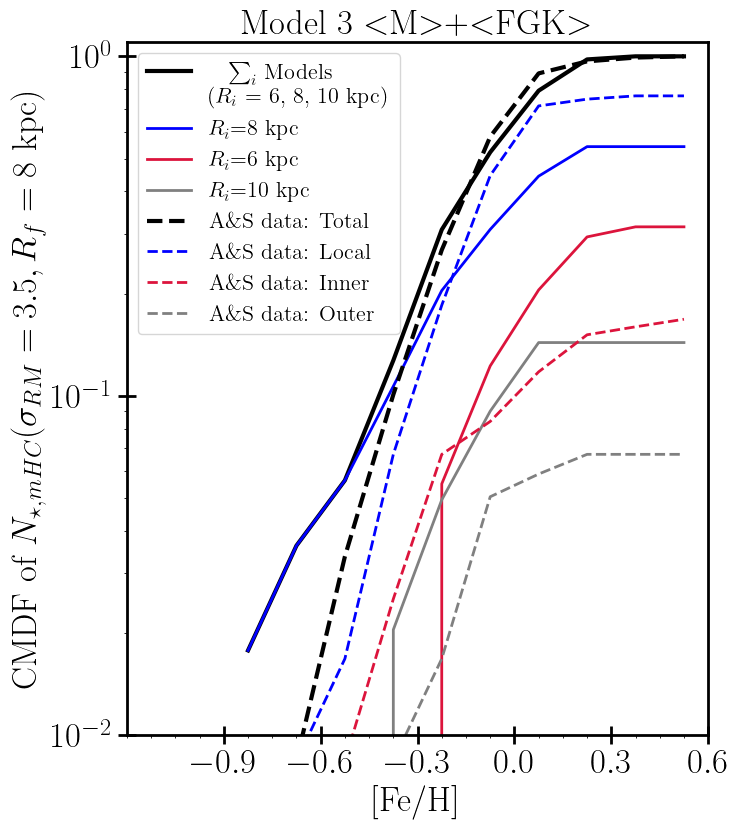}
      \includegraphics[scale=0.34]{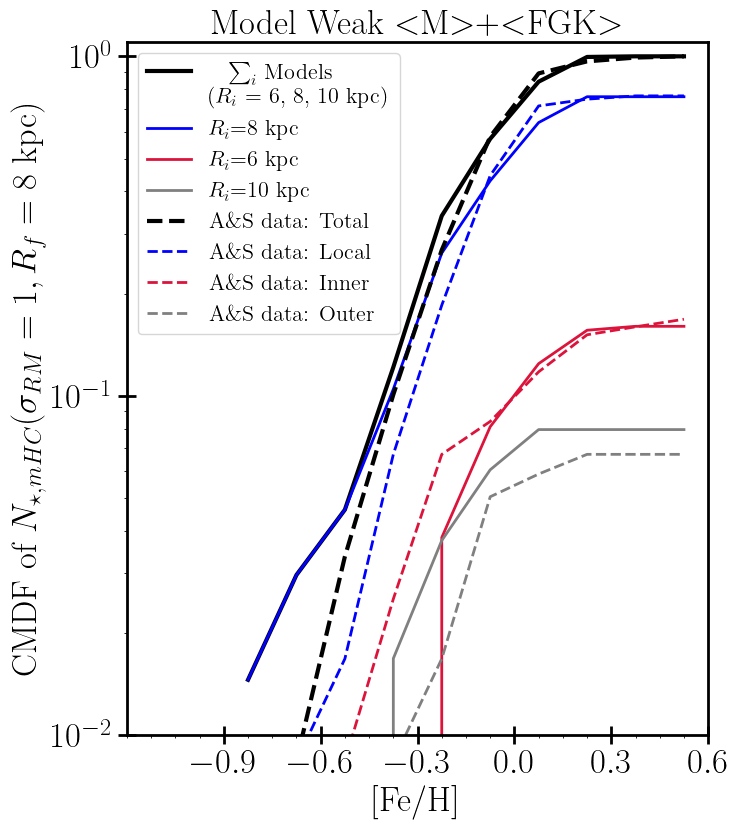}
     \caption{ CMDFs for FGK and M stars hosting habitable terrestrial planets ($N_{\star, mHC}$) in the solar vicinity predicted by models with stellar migration (solid lines).
     Model results are compared with the A$\&$S data sample  described  in Appendix \ref{app:data} for stars  hosting only low mass planets ($M_p<30M_{\oplus}$, dashed lines).  
     We  display the contribution from predicted stars that were born in situ within the solar annulus ($R_i = R_f = 8$ kpc,   solid blue lines), as well as those that migrated from  6 and 10 kpc (solid red and grey line, respectively)  to ensure consistency with the observed {\it local}, {\it inner}, and {\it outer} populations.  Model 3 ($\sigma_{RM}$=3.5) results are reported in the left panel. In the right one, we show the "Model Weak" adopting  a weaker radial
migration ($\sigma_{RM}$=1) while keeping all other model
parameters identical to those of Model 3. }
		\label{MDF_life_migr}
\end{figure*} 

 The predicted age-metallicity relation\footnote{To normalize the predicted chemical abundances to solar values, we adopt those proposed by \citet{lodders2009}.} shown in the right panel of Fig. \ref{SFR_ref} highlights that during the first Gyr of Galactic chemical evolution in the thick disc phase, the ISM undergoes a rapid metal enrichment. Notably, at an evolutionary time of  0.25 Gyr, the [Fe/H] = -1 dex threshold, identified by \citet{johnson_li2012} as the minimum metallicity required for the formation of terrestrial planets, is already reached. 
It is worth noting that the dilution phase characterizing the early evolution of the thin disc (at an evolutionary time of $\sim$ 3.25 Gyr) leads to a decrease in metallicity across all Galactic annular regions, with the effect being more pronounced at 18 kpc. However, this dilution phase does not reduce the [Fe/H] abundance below the -1 dex threshold, as shown in the lower panel of Fig. \ref{SFR_ref}. In conclusion, throughout the evolution of the Galactic disc at all radii — except for the first 0.25 Gyr — the metallicity remains above the threshold identified by \citet{johnson_li2012}. 

In Fig. \ref{multi_PE}, we show for the different Galactocentric distances the evolution of the probability $P_E$ for stars with different spectral types introduced by eq. (\ref{PE_bad}), where gas giants are assumed to be hazards for life.
It is worth noting that in the 3D space formed by  $P_E$, [Fe/H] and evolutionary
time, the transition between the thick and thin disc phases
emerges clearly. In fact, the stronger metal dilution in outer Galactic regions (see also Fig. \ref{SFR_ref}), together with the inside-out formation growth of the disc, implies a small effect of gas giants on the probability of finding Earth-like planets $P_E$. Conversely, the inner regions achieve lower values of $P_E$ due to the higher metallicities reached during the thin disc evolution phase.  This effect is more evident for retired A stars.

 Having in mind the evolution of the probability $P_E$ presented in Fig. \ref{multi_PE}, we compare in Fig. \ref{models_MFGKA} the number of stars hosting habitable Earth-like planets but not gas giant planets ($N_{\star,mHC}$) as a function of Galactic time and the Galactocentric distance, as predicted by Model 1 for M, FGK and  retired A stars, respectively. Consistently with the findings of \citet{spitoni2014,spitoni2017}, the distribution of  $N_{\star, mHC}$ over Galactic time and across Galactocentric distances reveals lower values in the innermost regions, primarily due to the high 
rates of SN events. Similarly, in the outskirts, the number remains low because of the reduced number of stars predicted by the model under the inside-out formation scenario.
Model 1, which assumes the case 2 SN damage scenario, shows the peak  of Galactic habitability  at 8 kpc for all spectral types.
The  numbers of M and FGK stars suitable for habitability ($N_{\star, mHC}$) are compatible with \citet{spitoni2017} predictions.   The main difference lies in the ratio between M and FGK stars, which is higher in this study due to the different adopted IMF
 (\citealt{Kroupa1993} versus \citealt{scalo1986}).
 Regarding retired A-type stars, we emphasise that the lifetime of the average stellar mass (weighted by the IMF, i.e. 1.78 \(M_{\sun}\) is  2.4 Gyr.
 This is the reason why in Fig. \ref{models_MFGKA} and in Fig.  \ref{A_GOOD} of Appendix \ref{app:B}, the distribution of stars hosting habitable terrestrial planets closely follows the star formation history, revealing two distinct clumps associated with the peak gas infall rates during the thick and thin disk phases. We underline that due to the short lifetimes of retired A-type stars, the development of complex life forms is unlikely on these objects.
In Fig. \ref{models1.2}, we show the evolution of the number of FGK stars hosting habitable Earth-like planets but not gas giant planets ($N_{\star, mHC}$) as a function of Galactic time and the Galactocentric distance predicted by  Model 2.
 In contrast to the results for Model 1, Model 2, which incorporates the Case 1 SN effect, predicts the peak habitability at 10 kpc. This difference arises because, as noted earlier, the average SN rate in the solar vicinity over the past 4.5 Gyr exceeds the predicted rates during the thin-disc phase for most of the Galaxy’s evolution at $R=8$ kpc. Importantly, at 4 kpc, no stars can host terrestrial planets capable of supporting life because of the too-high SN rates during Galactic history. Different thresholds in SN rate also impact the absolute values in the number of stars hosting terrestrial planets with conditions favourable for the development of life ($N_{\star, mHC}$). In Model 1,  the present-day 
$N_{\star, mHC}$ in the solar vicinity (annular region centered at 8 kpc from the Galactic center)
is $1.12 \times 10^9$ \footnote{We remind the reader that throughout this article, all predicted numbers of stars in different Galactic regions refer to annular regions 2 kpc wide, centered at the specified distances.}. At the same Galactocentric distance and time, Model 2 predicts $3.91 \times 10^8$ ($N_{\star, mHC}$), as highlighted in Fig. \ref{models1.2}.

In Fig. \ref{models_delay} of Appendix \ref{app:B}, we present the results for a “modified Model 1,” which accounts for the time necessary for a sustained increase in atmospheric O$_2$ to significant levels. On Earth, this process needed approximately 2.5 Gyr  \citep{Lyons2014}. Specifically, we considered delays of 1 Gyr and 3 Gyr in our analysis. At a distance of 8 kpc, we observe that the present-day total number of habitable stars, $N_{\star, mHC}$, decreases as the delay increases compared to the reference Model 1. For a 1 Gyr delay, $N_{\star, mHC}$ is $1.08 \times 10^9$, representing a 3.57\% decrease from the reference Model 1. For a 3 Gyr delay, $N_{\star, mHC}$ decreases further to $9.48 \times 10^8$, corresponding to a 21.00\% reduction.

\subsection{Model results with stellar migration}
\label{res_migr}

In this Section we present our findings in the presence of stellar migration.
In Fig. \ref{rirf8}, we show the temporal evolution of the total number of FGK stars with habitable terrestrial planets ($N_{\star, mHC}$) in the solar vicinity, highlighting the contributions from migrators originating in different Galactic regions. In the presence of stellar migration, we assume case 2 for the SN destruction scenario, adopting $\sigma_{RM}=3.5$ and $\sigma_{RM}=6$ (Model 3 and 4, see Table \ref{tab_A}). We stress that in our models, stellar migration is considered to begin with the onset of the thin disc phase, starting 3.25 Gyr after the beginning of Galactic evolution. 
Aside from the stars formed in situ, i.e. born at 8~kpc, the majority of FGK stars with habitable planets originate from 4, 6, and 10 kpc.
Looking in more detail, at the present time in Model 3 the total number of predicted stars with habitable terrestrial planets - currently residing in the solar vicinity - is $N_{\star, mHC}=1.14 \times 10^9$. Of these, 43.78\%  were formed in situ, 17.16\% originated from 4 kpc, 24.53\% from 6 kpc, and 11.57\% from 10 kpc. Similarly, in Model 4, the total number of stars hosting habitable planets is $1.01 \times 10^9$ $N_{\star, mHC}$. Of these, 38.82\% were formed in situ, 23.21\% originated from 4 kpc, 22.51\% from 6 kpc, and 10.27\% from 10 kpc. Concerning the total number of stars hosting life, we notice that the computed present-day value at 8 kpc  
is similar to that predicted by the model without stellar migration examined in the previous Section. This result can be explained by the fact that, in the solar vicinity, we reach a "balance" between stars leaving the solar vicinity and stars migrating toward this Galactic region (see e.g. \citealt{palla2022}). However, as we show later in the text, such a balance is not present ubiquitously across the Galactic disc.\\

\begin{figure}
	      \centering
          \includegraphics[scale=0.41]{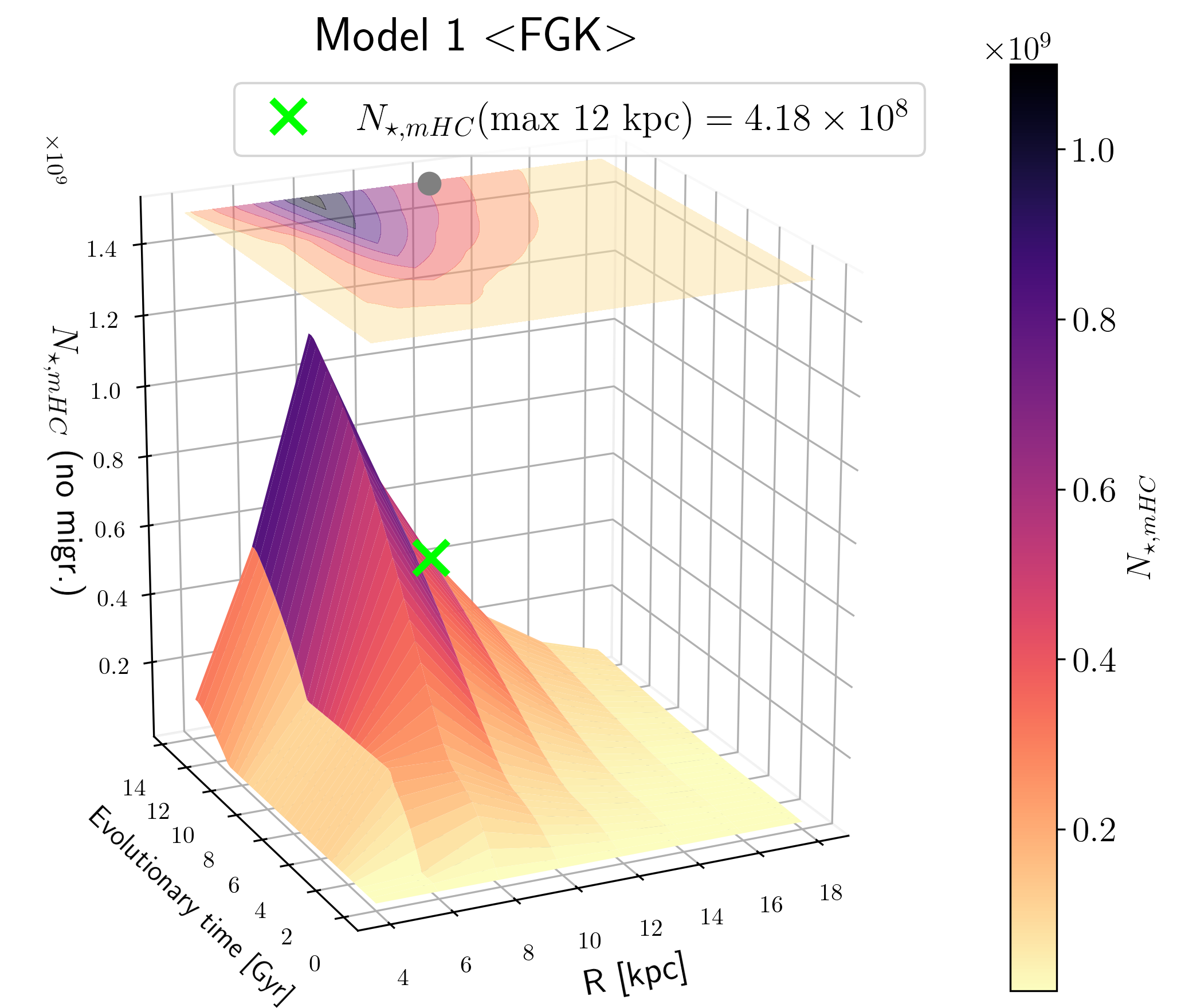}
         \includegraphics[scale=0.41]{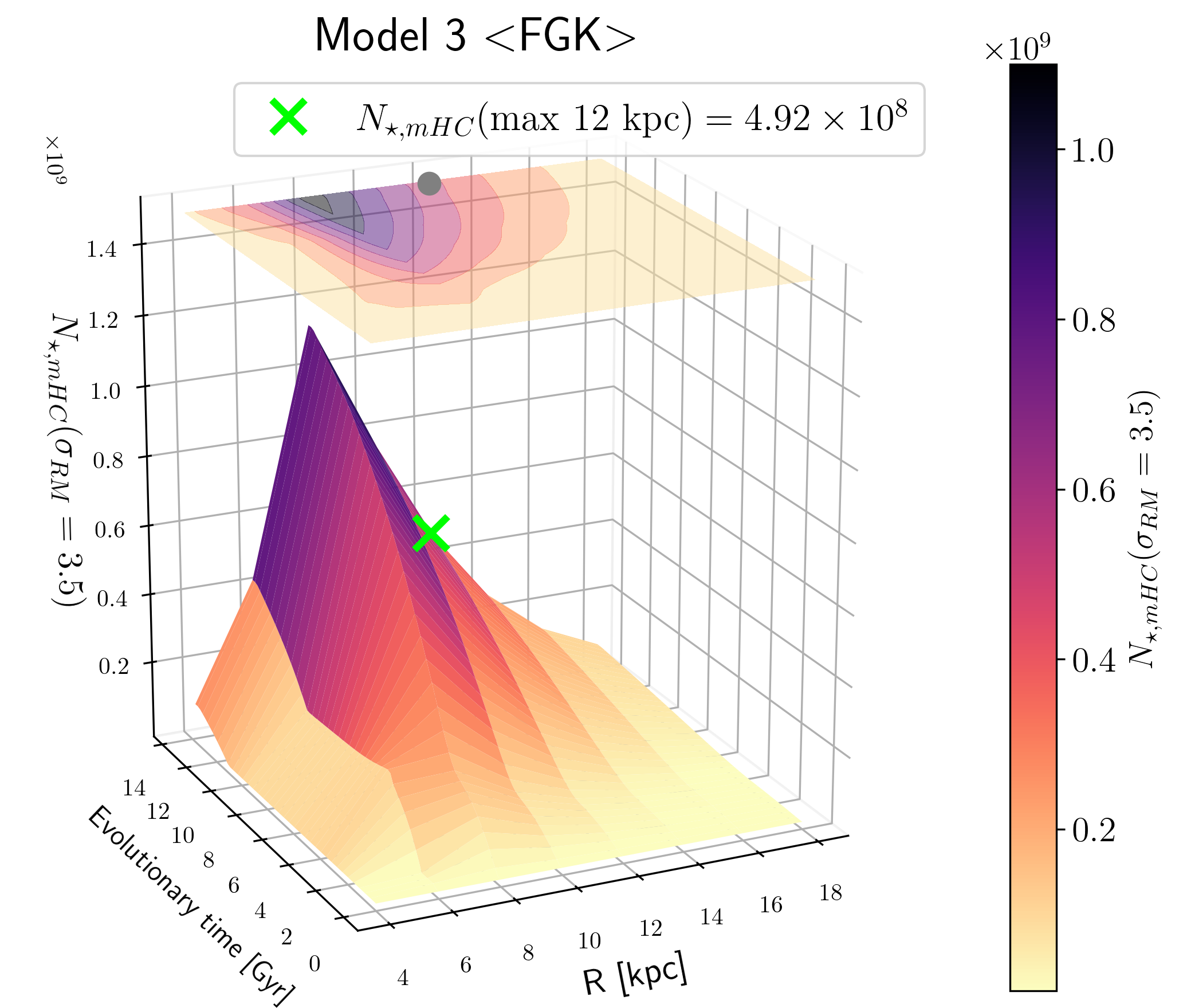}
      \includegraphics[scale=0.41]{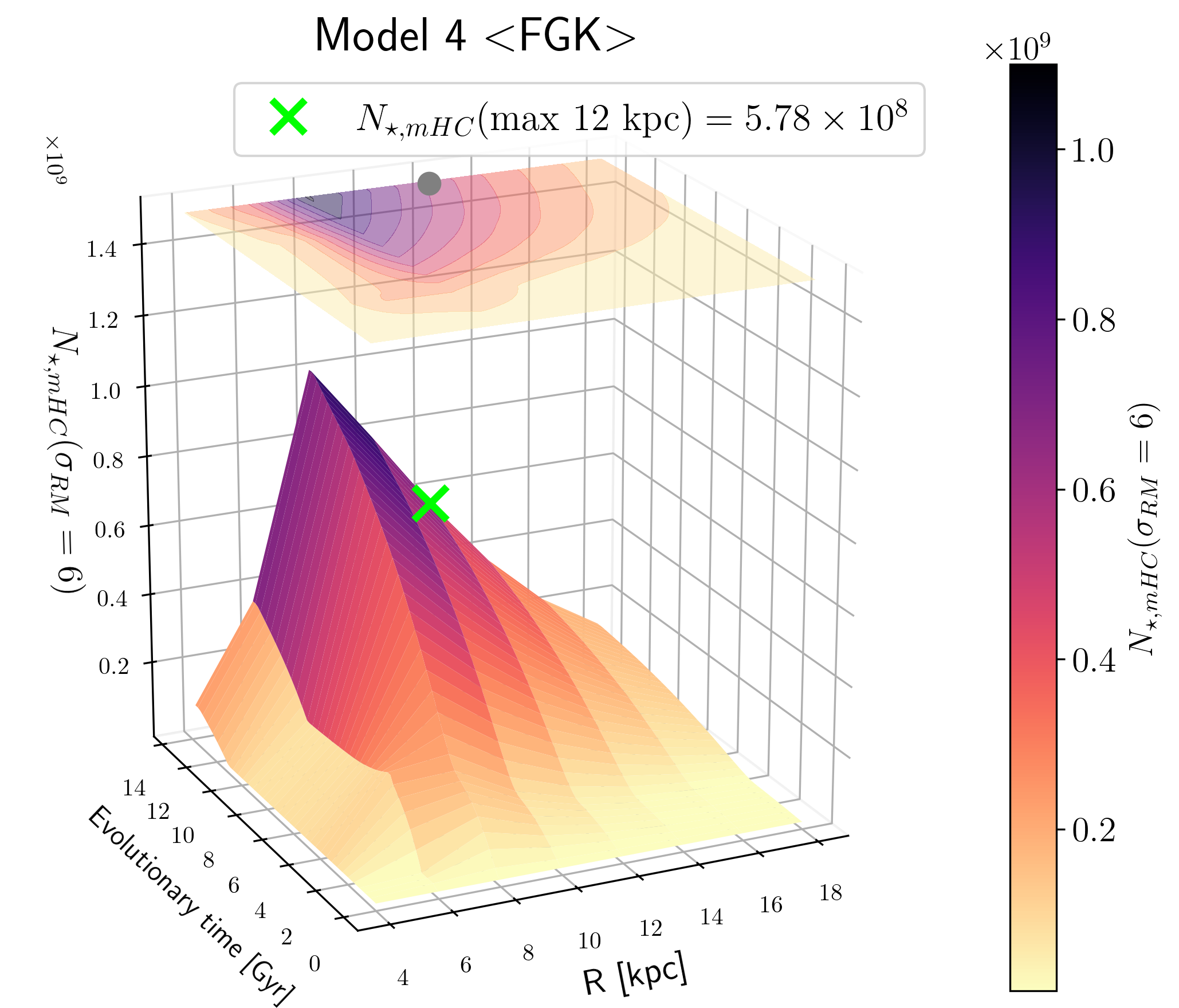}
  \caption{Total number of FGK stars  having Earths ($N_{\star, mHC}$) as a function of the Galactocentric distance and time  considering the Case 2 SN destruction scenario. 
  {\it Upper Panel}: Results from the reference model without including stellar migration (Model 1). 
  {\it Middle Panel}: Results from the model incorporating stellar migration as described in Section \ref{ss:migr_model}, with the radial migration strength fixed at $\sigma_{RM}=3.5$ kpc (Model 3). {\it Lower Panel}: Same as the middle panel but with a radial migration strength of $\sigma_{RM}=6$ kpc (Model 4).  In each panel, we highlight with a green cross the maximum number of FGK stars  hosting habitable terrestrial  planets predicted at 12 kpc (grey point in the 2D projection).}
 \label{life_migr}
\end{figure} 
\begin{figure}
	      \centering
\includegraphics[scale=0.42]{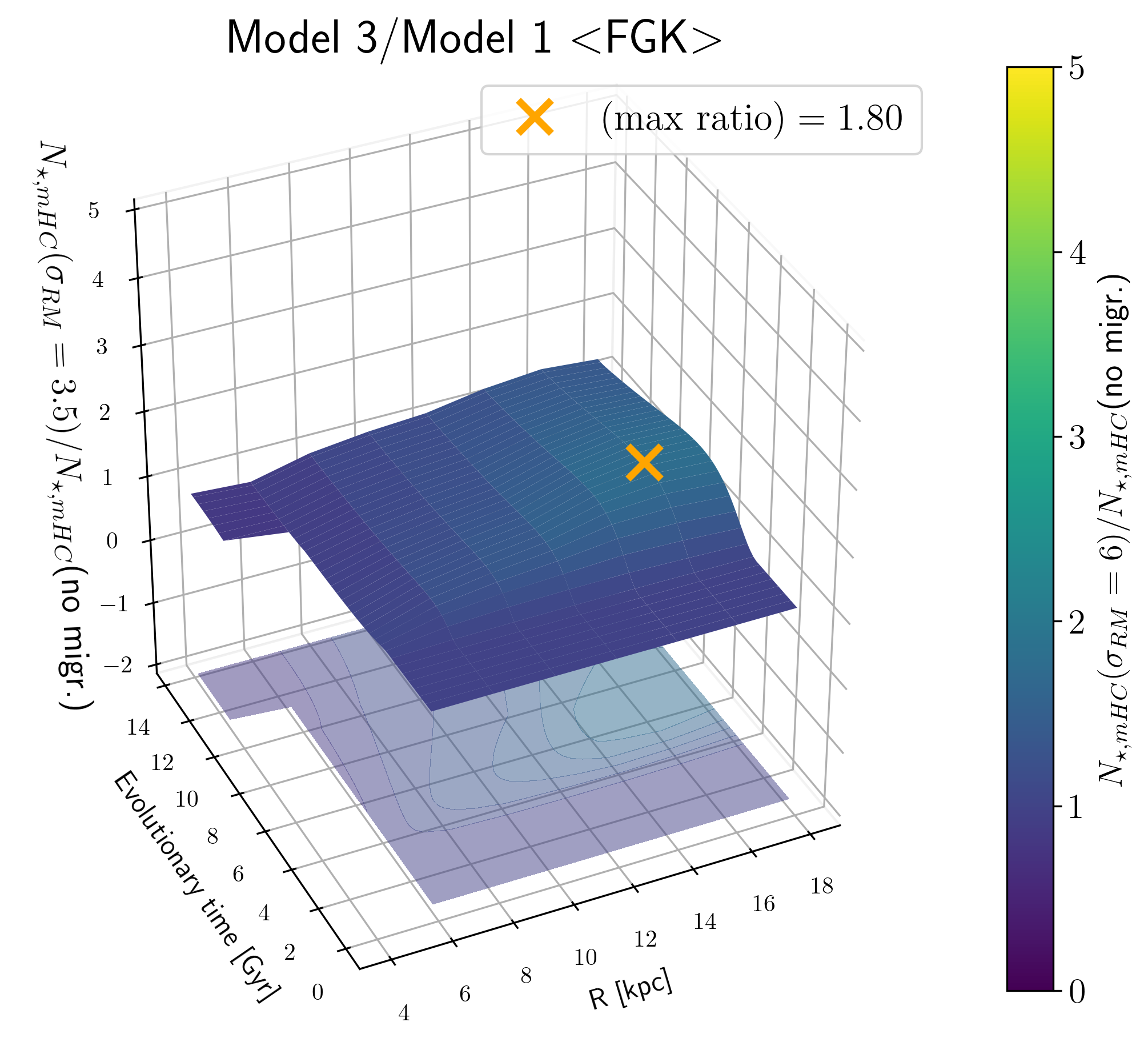}
     \includegraphics[scale=0.42]{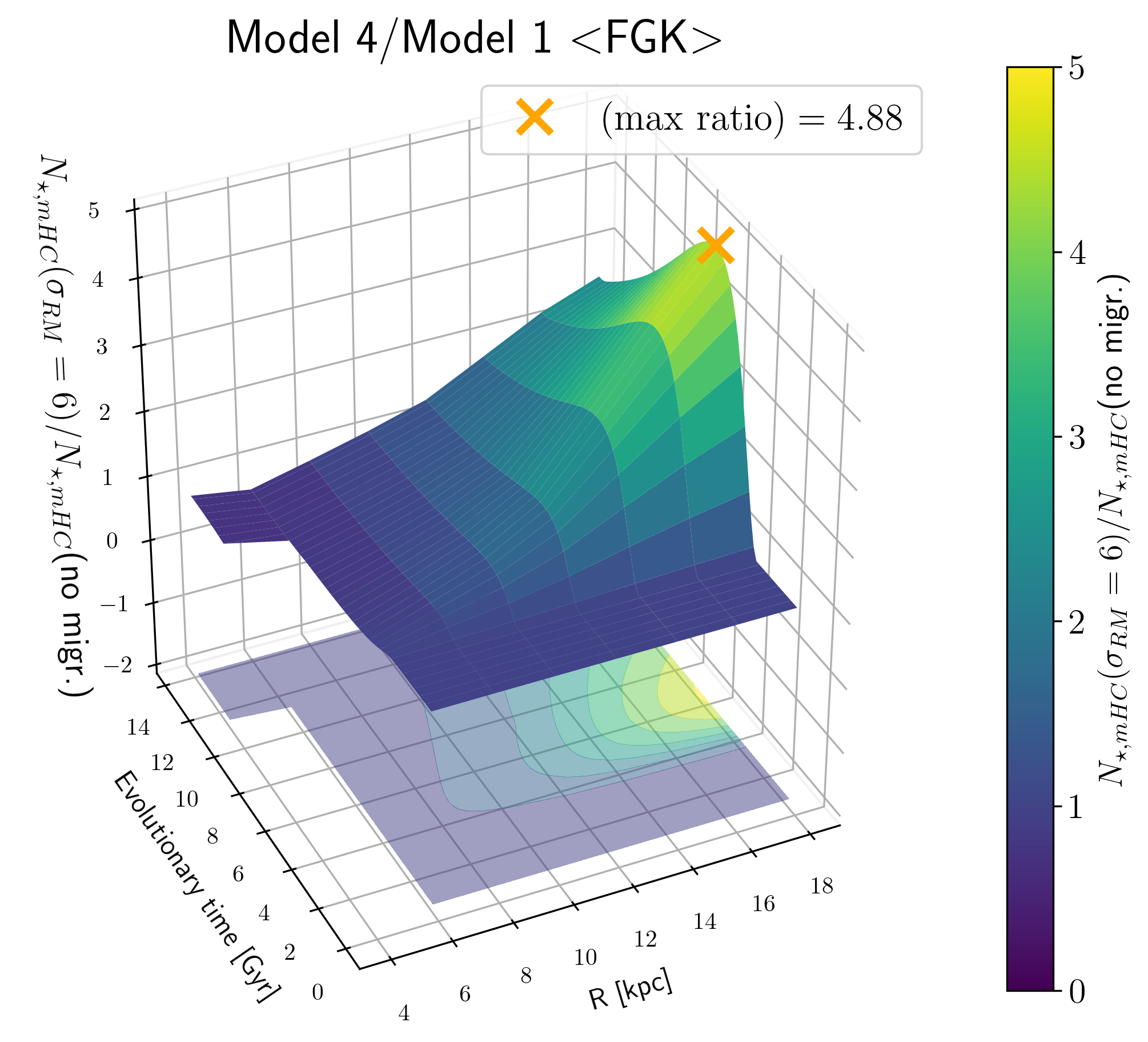}
  \caption{Ratios of the number of FGK stars hosting Earth-like planets ($N_{\star, mHC}$) in  models with stellar migration to the reference model without migration, as a function of Galactocentric distance and Galactic time. All considered models  adopt Case 2 for SN damage. {\it Upper panel}: stellar migration model with $\sigma_{RM} = 3.5$ (ratio of Model 3 to Model 1 in Table \ref{tab_A}). {\it Lower Panel}: stellar migration model with $\sigma_{RM} = 6$ (ratio of Model 4 to Model 1 in Table \ref{tab_A}).}
		\label{ratio}
\end{figure}

\begin{figure*}
          \centering
          \includegraphics[scale=0.34]{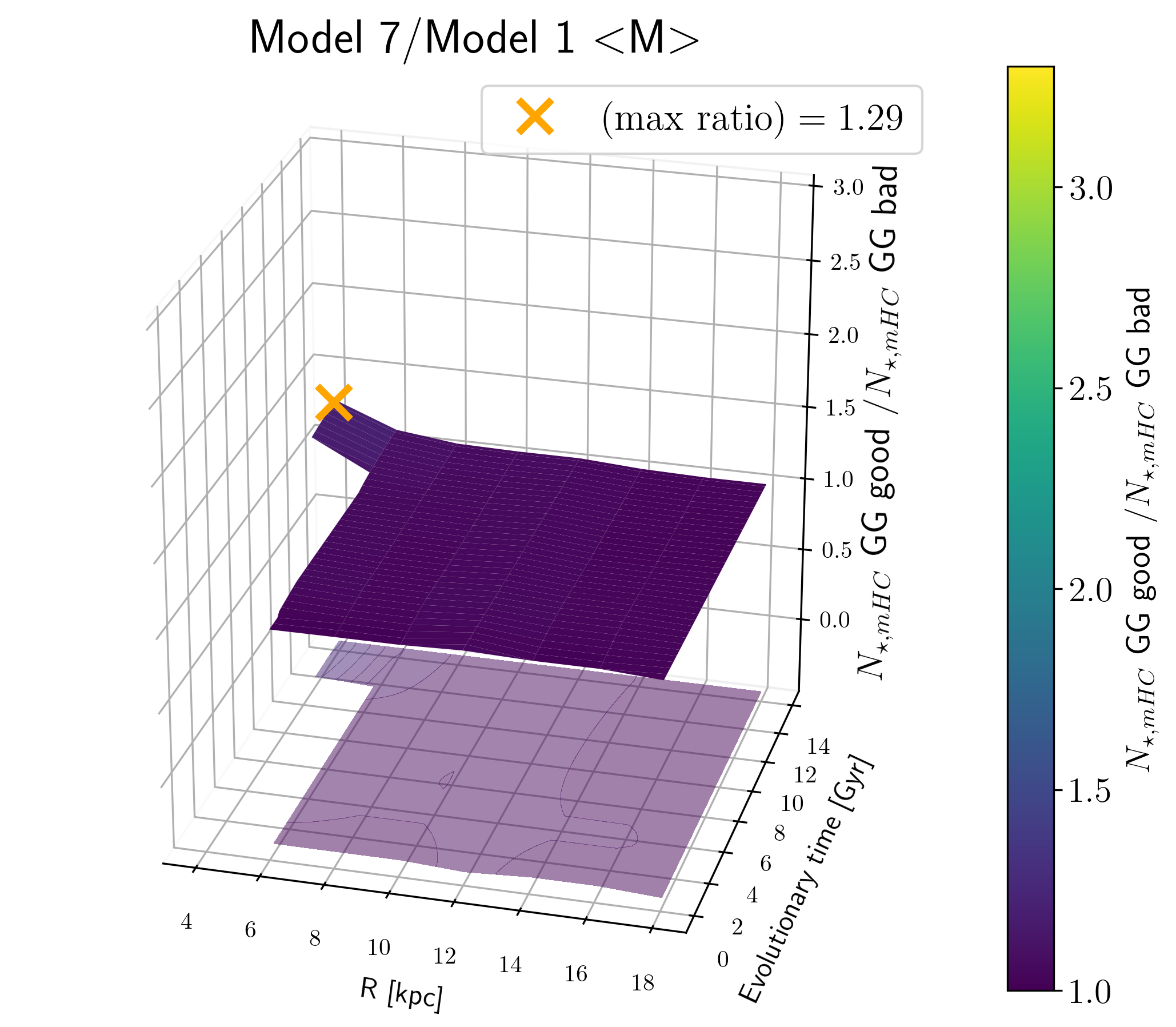}
     \includegraphics[scale=0.34]{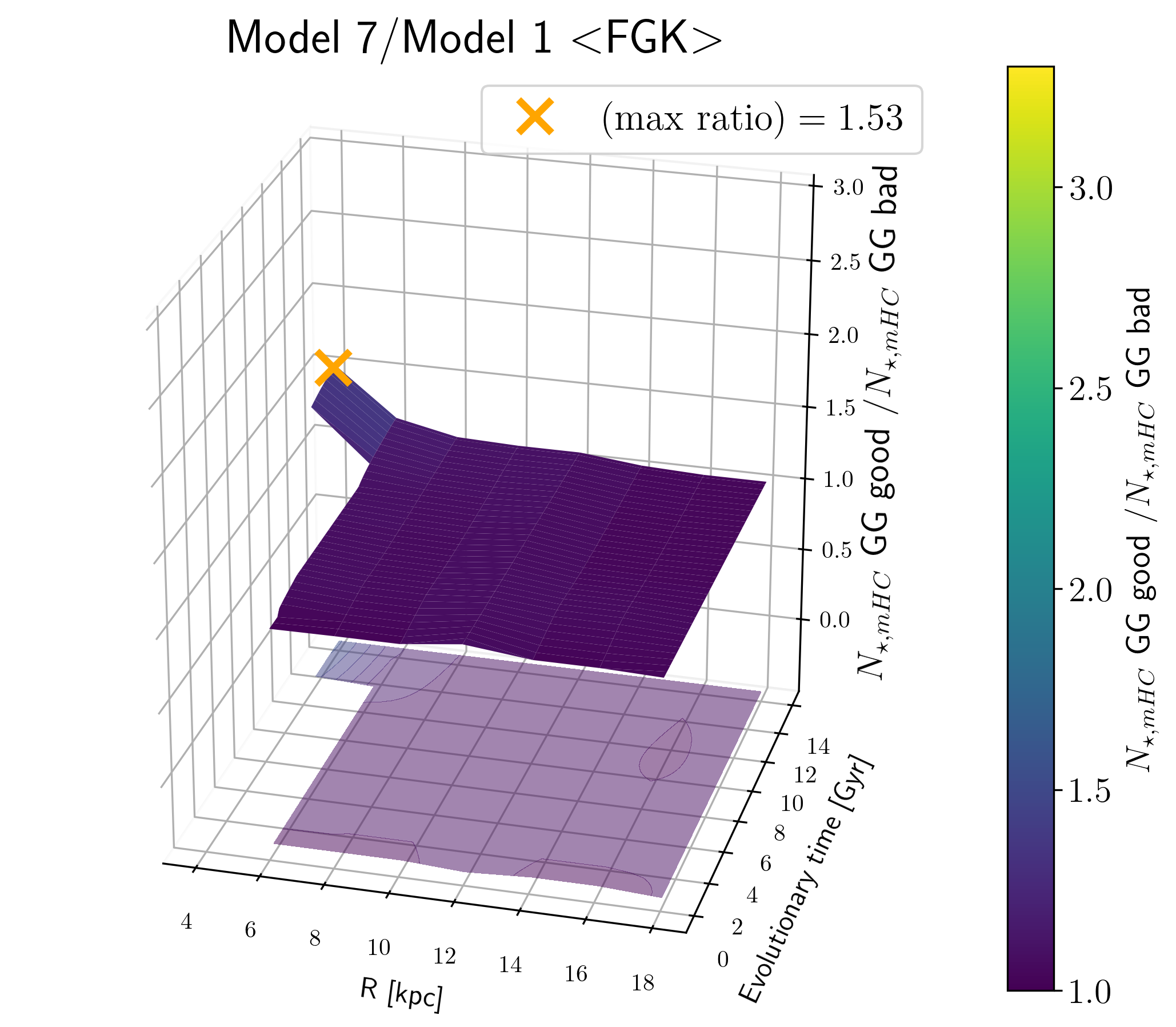}
\includegraphics[scale=0.34]{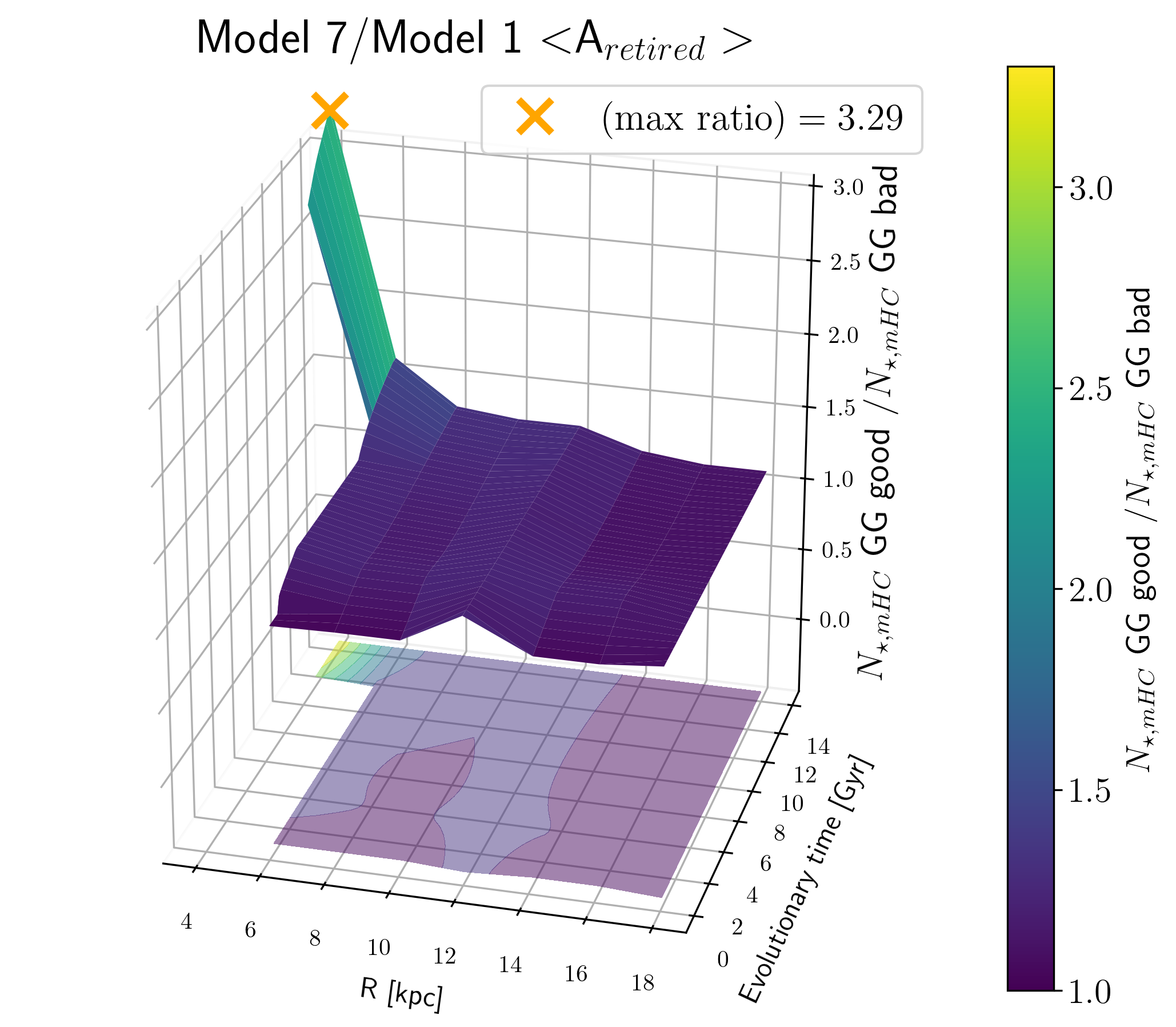}
 \includegraphics[scale=0.34]{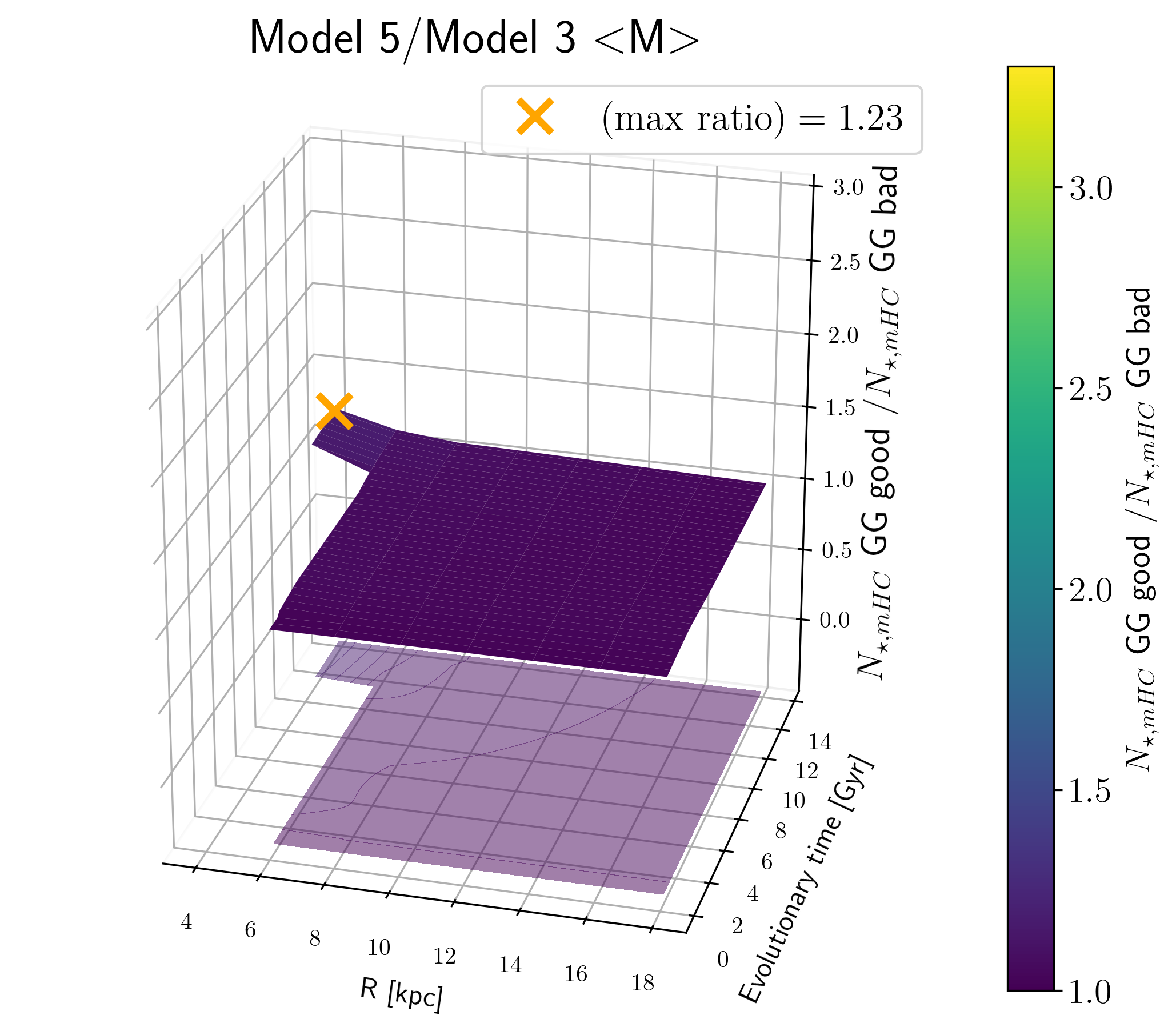}
\includegraphics[scale=0.34]{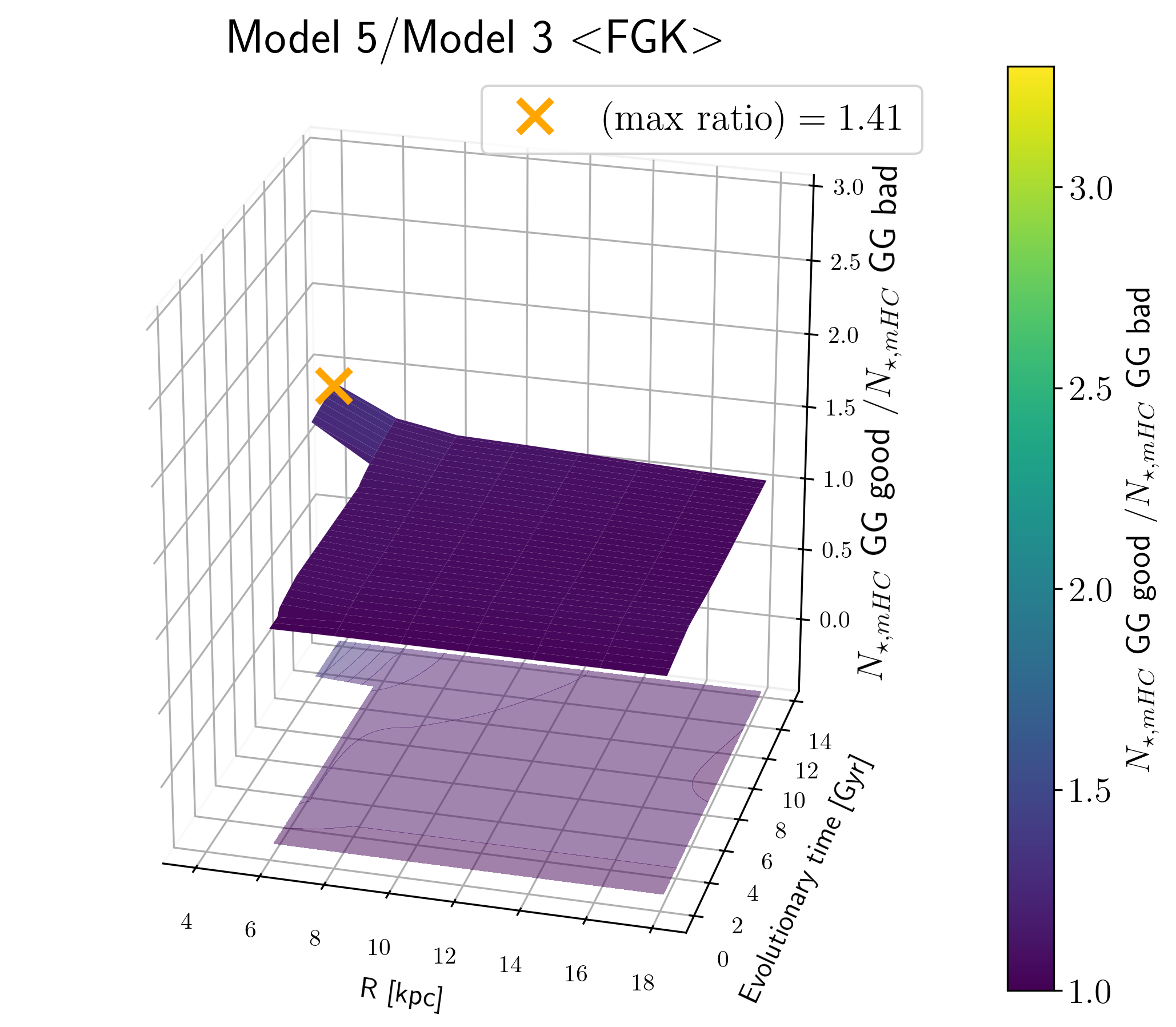}
\includegraphics[scale=0.34]{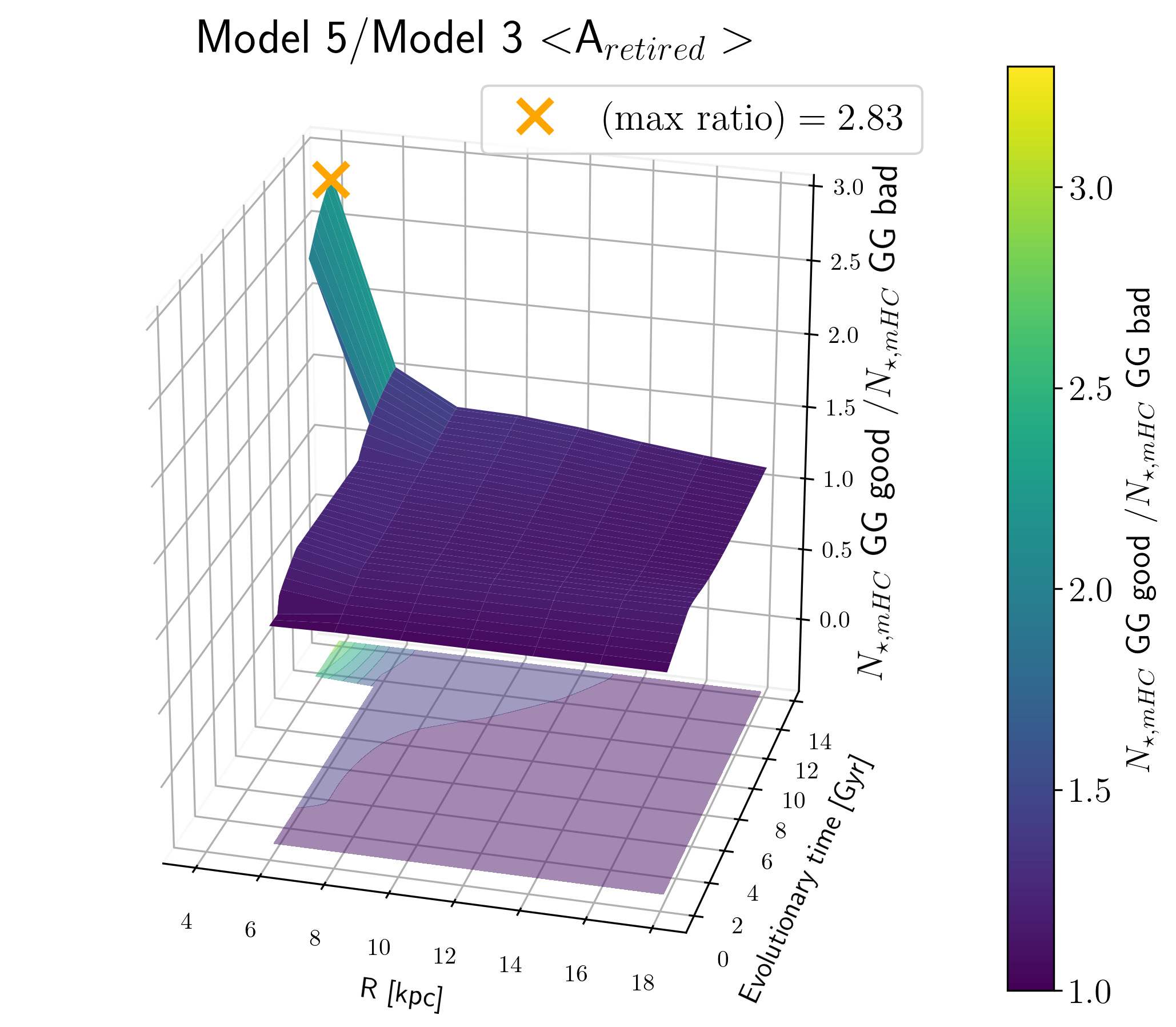}
     \caption{ Ratios between  the  predicted number of stars having Earths ($N_{\star, mHC}$)  of the two different prescriptions for $P_E$ as in Fig. \ref{GOOD}:  "GG GOOD" and "GG BAD" cases, respectively. Results for M, FGK and retired A are reported in the left, middle and right panels, respectively. Ratios are computed  as functions of the Galactocentric distance and Galactic time and all the displayed models consider the case 2 for the SN damage.  {\it Upper Panels}: reference models without stellar migration (ratio between Model 7 and Model 1 in Table \ref{tab_A}).
     {\it Lower   Panels}: models with stellar migration and $\sigma_{RM}=3.5 $  (ratio between Model 5 and Model 3 in Table \ref{tab_A}). In each panel the maximum value of the ratio is highlighted with an orange cross.
  }
		\label{ratio_GOOD}
\end{figure*} 

In the left panel of Fig. \ref{MDF_life_migr}, we present the cumulative metallicity distribution function (CMDF) predicted by Model 3 for  M and FGK stars possibly hosting life (\(N_{\star, mHC}\)) in the solar vicinity.
Our results are compared with the observed distribution of a sample of planet-hosting stars in the solar vicinity, comprising 124 stars (106 FGK-type and 18 M-type) that host planets with masses \( M_P < 30 M_{\oplus} \).
This sample is divided into {\it local} (born within 1 kpc of the Sun's location), {\it inner} (originating 1–2 kpc closer than the Sun to the Galactic center), and {\it outer} (located 1–2 kpc beyond the Sun) populations. This separation is based on the orbital properties of each star, from which we can extrapolate their origin in different regions of the Galactic disk.
Further details on the stellar sample are provided in the Appendix \ref{app:data}.
To ensure consistency with the observed {\it local}, {\it inner}, and {\it outer} populations (see Fig. \ref{fig:datamigration} in Appendix \ref{app:data}), we consider only predicted stars born at 6, 8, and 10 kpc in this analysis. 
The overall distribution of these populations is well reproduced by  Model 3 predictions, although the individual ones corresponding to stars originating from 6 and 10 kpc show that the model predicts a higher number of migrators. The data exhibit a higher contribution from the {\it local} sample (76.5\%) compared to Model 3 (54.2\%), indicating that a lower migration rate is needed to reproduce the data.
However, we must keep in mind that the sample used for comparison with the model accounts only for part of stellar migration. Indeed, data can account only for the "blurring" component (increase in the orbital eccentricity of a star, while maintaining the mean radius), while no direct observables are available for "churning" (change in mean radius / angular momentum of the orbit). Moreover, it is worth stressing that the sample is affected by observational biases, not representing all disc stars with minimal habitability conditions.
In the right panel of Fig. \ref{MDF_life_migr}, we explore a scenario with a weaker radial migration, adopting $\sigma_{RM}=1$ while keeping all other model parameters identical to those of Model 3 (`Model Weak'). In this case, the agreement with the observed sample across the different populations is nearly perfect. 
The reduction of the migration strength of a factor larger than 3 is actually in line with what was found in previous works (e.g. \citealt{Frankel2020}), where blurring accounts for a minor contribution in the radial migration process. However, we emphasize once again that a direct comparison with the observational sample is not entirely straightforward, as due to internal biases and unknown selection effects in the latter.

In Fig. \ref{life_migr}, we provide the full map of habitability in the 3D plot as functions of the Galactocentric distance and evolutionary times comparing models with and without migrations (Model 1, Model 3 and Model 4) for FGK stars. As discussed above, at 8 kpc the total number of stars is not dramatically affected by the migration. However, it is possible to note that the distribution of stars with habitability conditions in their hosting planets is wider towards external Galactocentric distances. For instance, at 12 kpc the present-day number of stars hosting habitable terrestrial planets 
increases from  $N_{\star, mHC}= 4.18 \times 10^8$ 
(Model 1, without migration) to $4.92 \times 10^8$ (Model 3, migration with $\sigma_{RM}=3.5$ kpc) and  $5.78 \times 10^8$ (Model 4, migration with $\sigma_{RM}=6$ kpc).

For this reason, in Fig. \ref{ratio} we present the ratio of $N_{\star, mHC}$ values predicted by models with stellar migration to those without, as a function of evolutionary time and Galactocentric distance. In this way, it is possible to better visualise the impact of stellar migration on the redistribution of stars hosting habitable planets.
This is particularly important in the outer Galactic regions, where the absolute value of the predicted number of stars is significantly smaller compared to the inner regions. In particular, Model 4 which is  characterized by $\sigma_{RM}=6$, shows a substantial increase in the number of stars hosting habitable planets in the outer Galactic regions. This occurs because stellar migration enables stars to move towards more external parts, characterised by  lower star formation activity but reduced SN rates (see Fig. \ref{SFR_ref}), creating more favourable conditions for life to develop. For example, in Model 4, at a Galactocentric distance of 18 kpc, the number of habitable planets increases by a factor of approximately 4.9 at around $t=6.1$ Gyr of Galactic evolution for FGK stars \footnote{This predicted ratio is approximately 4.9  for also both M stars and retired A stars.}. It is important to note that at 4 kpc, ratio values are available only for the last 2 Gyr of Galactic evolution, during which the model without stellar migration exhibits a SN rate below the threshold defined by the Case 2 SN destruction prescriptions (see Fig. \ref{SFR_ref}).
The increased number of stars hosting habitable planets in the outer Galaxy due to stellar migration widens the outer boundaries of the GHZ.
Such a result is particularly relevant considering that the abundances of organic molecules found in outer Galaxy star-forming regions are comparable to those measured near the Sun \citep[e.g.][]{bernal21,fontani22,fontani24}, despite the sub-Solar metallicity.
Some of these molecules may have been the seeds of organic compounds that favoured the emergence of life in the Solar System and elsewhere in the Universe.
Therefore, both findings go in the same direction that the conditions in the outer Galaxy are likely more favourable to host habitable planets than we previously thought. 

 As discussed in Section \ref{molero}, our analysis adopts a delay of $t_{\mathrm{max}} = 3.25$ Gyr between the two gas infall episodes, in order to match the APOGEE observational constraints \citep{Palla2020}. To assess the sensitivity of our results to this parameter, we also explored models with $t_{\mathrm{max}} = 1$ and $5$ Gyr. Relative to Model 4 (see lower panel of Fig.~\ref{life_migr}), we find that at the Galactocentric distance of 12 kpc the number of FGK stars with minimum habitability conditions ($N_{\star, \mathrm{mHC}}$) increases by approximately 6.4\% for $t_{\mathrm{max}} = 1$ Gyr, and decreases by about 8.0\% for $t_{\mathrm{max}} = 5$ Gyr. 
This behaviour reflects our assumption - following the prescriptions of \citet{frankel2018} - that stellar migration occurs only during the thin-disc phase, which is longer in the $t_{\mathrm{max}} = 1$ Gyr model and shorter in the $t_{\mathrm{max}} = 5$ Gyr case.

\subsection{Results with gas giants as catalysts for terrestrial planets}
\label{GG_GOOD_sect}

As anticipated in Section \ref{sect_model_good},  
another main purpose of this work is to test the  scenario  where gas giants  act as catalysts for terrestrial planet formation using the new probability $P_E$  "GG GOOD", introduced  in eq. (\ref{PE_good}) and lower panel of Fig.   \ref{GOOD}. 
Figure \ref{ratio_GOOD} shows the ratios of the predicted number of M, FGK and retired A stars hosting habitable Earth-like planets ($N_{\star, mHC}$) under two distinct $P_E$ scenarios, “GG GOOD” and “GG BAD,” as defined in Fig. \ref{GOOD}. Specifically, the following cases are analysed:
i) No migration (ratio between Model 7 and Model 1 from Table \ref{tab_A});
ii) Stellar migration with $\sigma_{RM} = 3.5$ (the ratio between Model 5 and Model 3 from Table \ref{tab_A}).
As noted in Section \ref{sect_model_good}, the two $P_E$ probabilities, “GG GOOD” and “GG BAD”, exhibit significant differences only at super-solar metallicities. 
Consequently, we expect disparities primarily in the innermost regions of the Galactic disc, where higher metallicities are predicted. 
For all the stellar spectral types considered, the differences between the GHZ maps in the “GG GOOD” and “GG BAD” scenarios, as shown in Fig. \ref{ratio_GOOD}, are primarily concentrated in the annular region centered at 4 kpc. As expected from Fig. \ref{GOOD}, the highest GG GOOD/GG BAD ratio is found for retired A stars, reaching 3.3 in the model without stellar migration. For FGK stars, the maximum ratio is 1.53, while for M stars - where $P_E$ remains nearly constant across all metallicity ranges in both scenarios - it is 1.29.
In Fig. \ref{A_GOOD} of Appendix \ref{app:B}, we show the total number of retired A stars  having Earths ($N_{\star, mHC}$) as a function of the Galactocentric distance considering  different gas giant effects. It is possible to appreciate that at 8 kpc the number of stars within the GG GOOD scenario increased by a factor of 1.23.
However, these factors decrease when stellar migration is included, as illustrated in the lower panels of Fig. \ref{ratio_GOOD}.
This reduction is a natural consequence of the net effect of stellar migration, which - on average - redistributes stars from the innermost regions towards outer Galactic areas.
In general, the GG GOOD and GG BAD scenarios yield minimal differences in  numbers of the predicted stars hosting habitable planets ($N_{\star \, mHC}$). In fact, regardless of the presence of stellar migration,  the enhancement factor is practically negligible in the regions where the peak of habitability occurs (i.e. in the solar vicinity for the Case 2 SN damage model).

\section{Conclusions}
\label{conclusions}

In this paper, we presented new GHZ  maps  for the Milky Way disc constructed by means of detailed and well-tested chemical evolution models (see \citealt{molero2023,molero2025}) taking into account, for the first time, the effects of the stellar migration, following the parametric approach suggested by \citet{frankel2018} and  \citet{palla2022}. 
Moreover,  we explored  different scenarios for the effect of the presence of gas giant planets on the formation of terrestrials, considering both positive and negative effects. 
The main results can be summarized as follows:

\begin{itemize}

\item The Galactic habitability maps are highly sensitive to the adopted threshold for SN rates required for destruction to be effective. Model 1, which assumes the highest SN threshold (case 2), shows the peak of Galactic habitability at 8 kpc. In contrast, Model 2, which adopts the case 1 SN effect, predicts the peak shifted at 10 kpc.
\item   
At the present time, in the solar vicinity, the total number of predicted stars with habitable terrestrial planets, $N_{\star, mHC}$, is similar in models with or without stellar migration. However,  stellar migration leads to a redistribution of stars and a large fraction of them originated in different Galactic regions.
For instance, in Model 3 (where the migration strength is fixed at the value of $\sigma_{RM}=3.5$ kpc, as suggested by chemical evolution studies) 17.16\% of FGK stars hosting habitable planets are born at 4 kpc, 24.53\% at 6 kpc, 11.57\% at 10 kpc, and 43.78\% is formed in situ.

\item Stellar migration has a larger impact in the outskirts of the  Galactic disc. In the annular region centered at 18 kpc, the number  $N_{\star, mHC}$ of stars hosting habitable planets is increased by a factor of $\sim5$ compared to the reference model without stellar migration when the extreme migration case is considered (Model 4).

\item  The hypothesis that gas giant planets facilitate the formation of terrestrial planets  has the most pronounced effect in the inner Galactic disc (in the annular region centered at 4 kpc from the Galactic center). However, we find that this increased probability  is also mitigated by stellar migration.  In particular, at the present time, when stellar migration is taken into
account, the number of FGK stars hosting  habitable terrestrial planets  in the inner ring centered at 4 kpc
is approximately 1.4 times higher than in scenarios where gas giants are assumed to hinder the formation and evolution of Earth-like
planets. Without stellar migration, this factor increases to 1.5. Even larger ratios are predicted for terrestrial planets orbiting retired
A stars, reaching 2.8 in models with stellar migration and 3.3 in models without it. 
In general, the two scenarios (i.e. GG BAD and GG GOOD cases in Table \ref{tab_A}) yield to
minimal differences in absolute numbers of the predicted stars hosting habitable planets ($N_{\star \, mHC}$). In fact, regardless of the presence of stellar migration,  the enhancement factor is practically negligible in the regions where the peak of habitability occurs for M and FGK stars (i.e. in the solar vicinity).
\end{itemize}
In conclusion, in this study, we have significantly expanded the exploration of the parameter space defining the Galactic Habitable Zone, compared to previous analyses present in literature \citep[e.g.][]{lineweaver2004, prantzos2008, spitoni2017}. 
 Our findings are particularly relevant in the context of upcoming space missions, such as the ESA PLAnetary Transits and Oscillations of Stars (PLATO; \citealt{rauer2024}), the ESA Ariel space mission \citep{tinetti2018,tinetti2022} and Large Interferometer For Exoplanets (LIFE, \citealt{quanz2022}). 
  These missions will deliver unprecedented data on planetary properties, orbital architectures, and atmospheric compositions. To fully interpret this information, it will be essential to place exoplanets within a broader Galactic framework that accounts for stellar chemical composition, formation environment, and radial migration.
This work contributes to the foundational understanding of planetary formation and habitability from circumstellar to Galactic scales, forming part of the necessary groundwork to interpret the influx of data expected from upcoming discoveries.

\begin{acknowledgements}
The authors thank the referee, R. Heller, for helpful suggestions that improved the quality of the manuscript.
The analysis and work presented in this paper were developed during and after discussions at the ``Molecules and planets in the outer Galaxy: is there a boundary of the Galactic Habitable Zone?'' workshop, funded by INAF.
E.S., F.M. and G.C.  thank I.N.A.F. for the  
1.05.24.07.02 Mini Grant - LEGARE "Linking the chemical Evolution of Galactic discs AcRoss diversE scales: from the thin disc to the nuclear stellar disc" (PI E. Spitoni).
E.S., L.M., M.T. and G.C.  thank I.N.A.F. for the 1.05.23.01.09 Large Grant - Beyond metallicity: Exploiting the full POtential of CHemical elements (EPOCH) (ref. Laura Magrini).
M.P. and D.R. acknowledge financial support from the project "LEGO – Reconstructing the building blocks of the Galaxy by chemical tagging" granted by the Italian MUR through contract PRIN2022LLP8TK\_001 (PI: A.~Mucciarelli). D.R. acknowledges I.N.A.F. for financial support through the 1.05.12.06.08 Theory Grant - An in-depth theoretical study of CNO element evolution in galaxies (ref. Donatella Romano). 
F.M. thanks I.N.A.F. for the 1.05.12.06.05 Theory Grant - Galactic archaeology with radioactive and stable nuclei.  F.M. thanks also for support from Project PRIN MUR 2022 (code 2022ARWP9C) "Early Formation and Evolution of Bulge and HalO (EFEBHO)" (PI: M. Marconi), M.M. thanks for the support from the Deutsche Forschungsgemeinschaft (DFG, German Research Foundation) – Project-ID 279384907 – SFB 1245, the State of Hessen within the Research Cluster ELEMENTS (Project ID 500/10.006).
L.M. and G.C. thank INAF for the support with the Mini-Grants Checs  (1.05.23.04.02), and the financial support under the National Recovery and Resilience Plan (NRRP), Mission 4, Component 2, Investment 1.1, Call for tender No. 104 published on 2.2.2022 by the Italian Ministry of University and Research (MUR), funded by the European Union – NextGenerationEU – Project ‘Cosmic POT’ Grant Assignment Decree No. 2022X4TM3H (PI: L. Magrini) by the Italian Ministry of the University and Research (MUR).
C.D. acknowledges financial support from the INAF initiative ``IAF Astronomy Fellowships in Italy'', grant name \textit{GExoLife}. 
This work used the Ariel Stellar Catalogue developed by the Stellar Characterisation WG in preparation of the ESA Ariel space mission. The catalogue can be found at  \url{bit.ly/ArielStellarCatalogue}.
\end{acknowledgements}

\bibliographystyle{aa} 
\bibliography{disk}

\begin{appendix}  

\section{The chemical evolution model}  
\label{app:A}  
\subsection{Nucleosynthesis prescriptions} 
\label{app:nucleo} 
The nucleosynthesis prescriptions for the reference chemical evolution model adopted in this work and described in  Section \ref{molero}  follow those of the MWG11 model from \citet{romano2019}. For low- and intermediate-mass stars, we adopt the stellar yields from \citet{ventura2013}.
 Our models also account for the nucleosynthetic outcomes of binary systems that give rise to Type Ia SNe, assuming the single-degenerate scenario for progenitors as described by \citet{matteucci2001} and references therein, with nucleosynthetic yields from \citet{iwamoto1999}.
The yield for massive stars in the presence of stellar rotation are the ones of \citet{limongi2018}. A variable initial rotational velocity for massive stars is considered, favouring high rotational speeds at low metallicities while assuming no rotation at solar metallicity (see also \citealp{prantzos2018, rizzuti2019, molero2024}). Specifically, we adopt $v_{\text{rot}} = 300 \, \text{km s}^{-1} \, \text{for} \, [\text{Fe}/\text{H}]$$ < -1, \, v_{\text{rot}} = 0 \, \text{for} \, [\text{Fe}/\text{H}] \geq -1$ dex. 
\subsection{Other model predictions} 
\begin{figure}
	      \centering
     \includegraphics[scale=0.4]{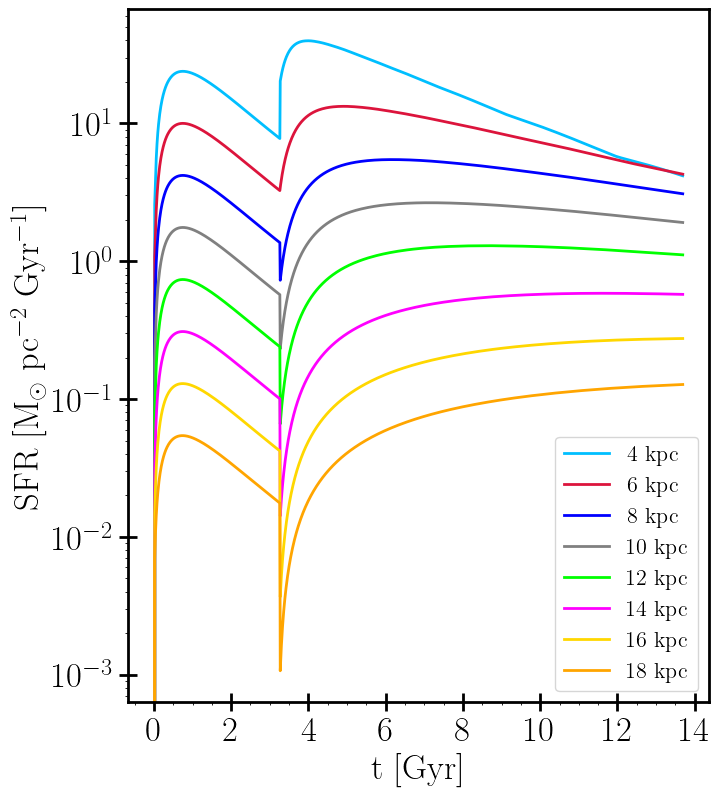}
  \caption{ Predicted time evolution of the SFR of  our multi-zone chemical evolution model presented in Section \ref{sec_results_nomigr} as a function of the evolutionary time $t$ and the Galactocentric distance.  
}
		\label{SFR_app}
\end{figure}

\begin{figure}
	      \centering
     \includegraphics[scale=0.43]{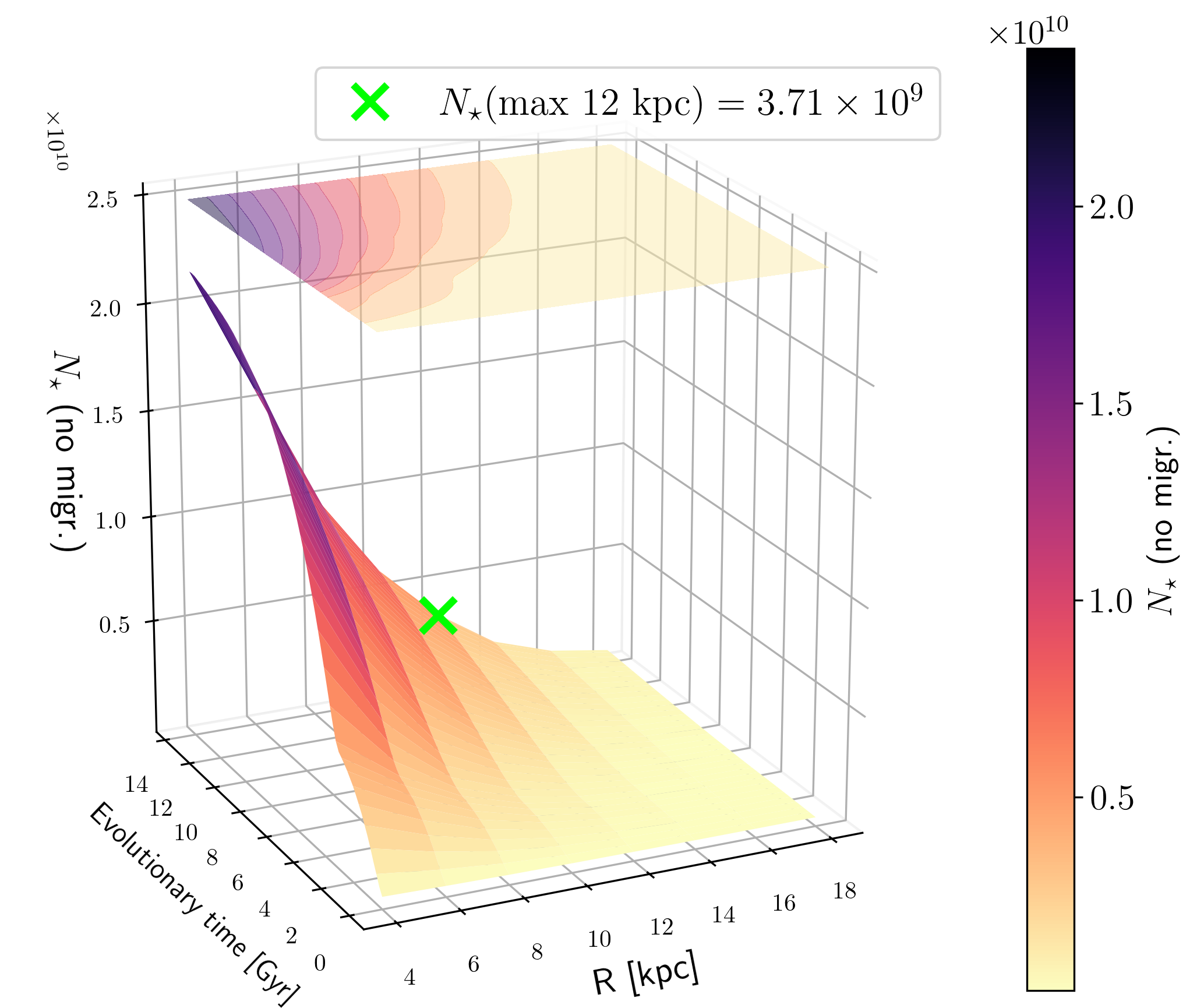}
      \includegraphics[scale=0.43]{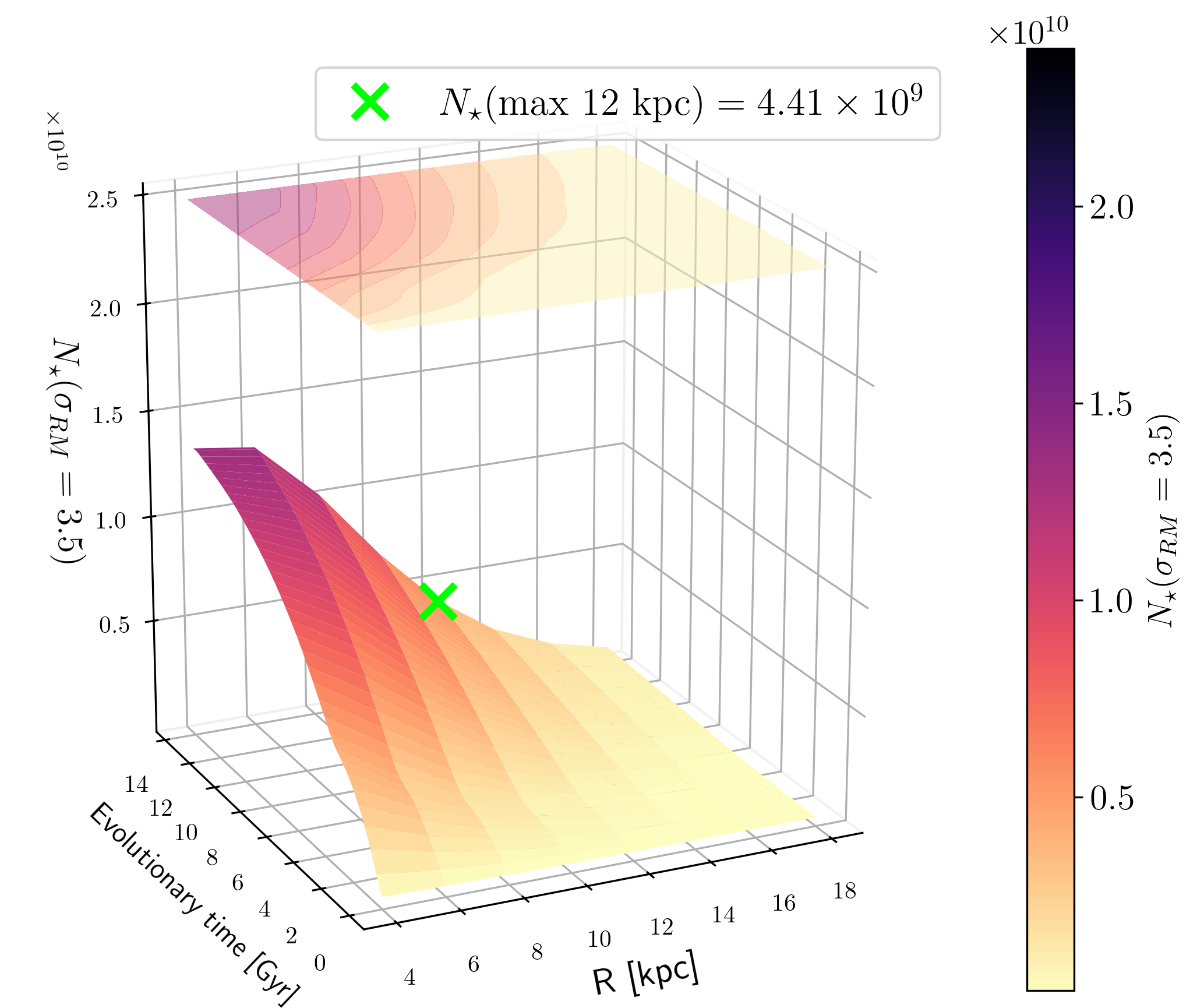}
       \includegraphics[scale=0.43]{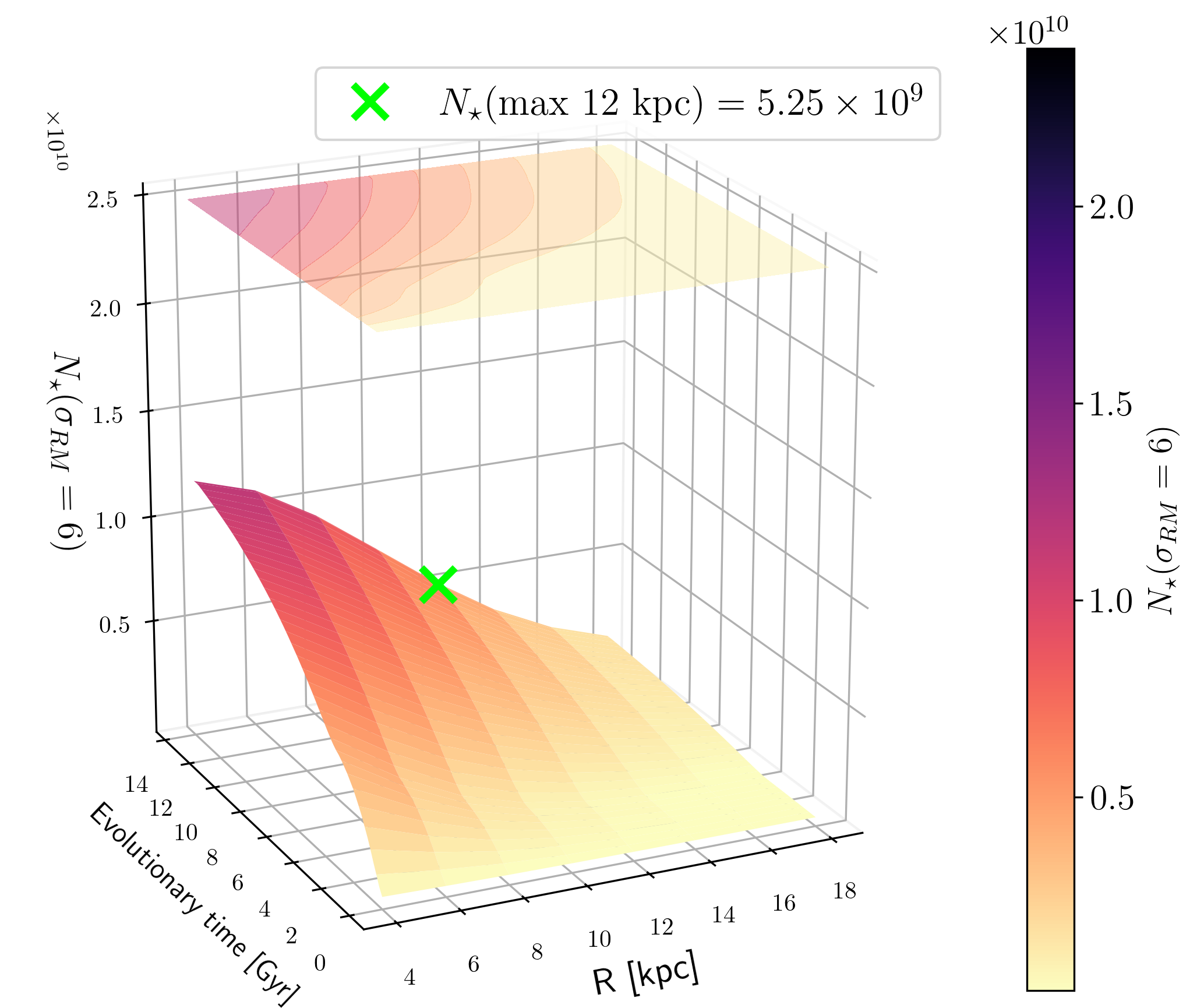}
  \caption{ The total number of predicted stars ($N_{\star}$)  as a function of the Galactocentric distance and the Galactic time. 
     {\it Upper Panel}: Results from the reference model without including stellar migration. {\it Middle Panel}: Results from the model incorporating stellar migration as described in Section \ref{ss:migr_model}, with the radial migration strength fixed at $\sigma_{RM}=3.5$. {\it Lower Panel}: Same as the middle panel but with a radial migration strength of $\sigma_{RM}=6$.  In each panel, we highlight with a green cross the maximum number of stars computed at 12 kpc. }
		\label{total}
\end{figure}

In Fig. \ref{SFR_app}, we show the predicted time evolution of the SFR of  our multi-zone chemical evolution model presented in Section \ref{molero}.
In Fig. \ref{total}, we compare the total number of stars predicted by the reference model without stellar migration to those where radial migration is taken into account. For the latter, we fix the migration strength $\sigma_{RM}$ to two values: 3.5 kpc, as used in \citet{palla2022} and \citet{frankel2018}, and 6 kpc, representing an extreme case of stellar radial migration.
As expected, the stellar migration has the effect of increasing the number of stars in the outer parts of the Galactic disc. For instance, at a Galactocentric distance of 12 kpc, the present day total number of stars predicted by the reference model without migration is  $3.71 \times 10^9$. When migration is considered with $\sigma_{RM}=3.5$, the number increases to $4.41 \times 10^9$, and in the extreme case with $\sigma_{RM}=6$, the total reaches $5.25 \times 10^9$.

\section{Other GHZ results}  
\label{app:B}
\begin{figure}
          \centering
         \includegraphics[scale=0.52]{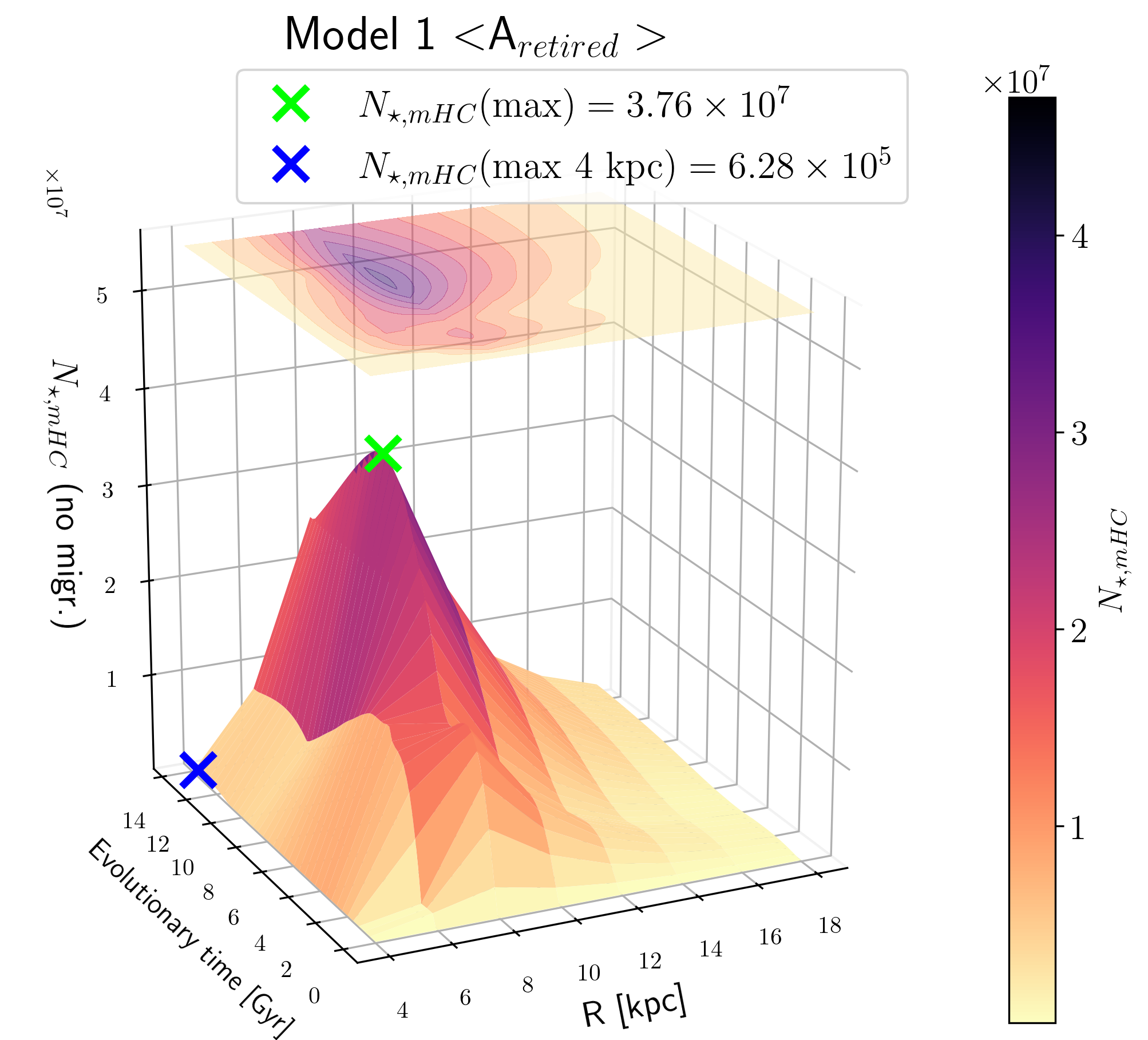}
\includegraphics[scale=0.52]{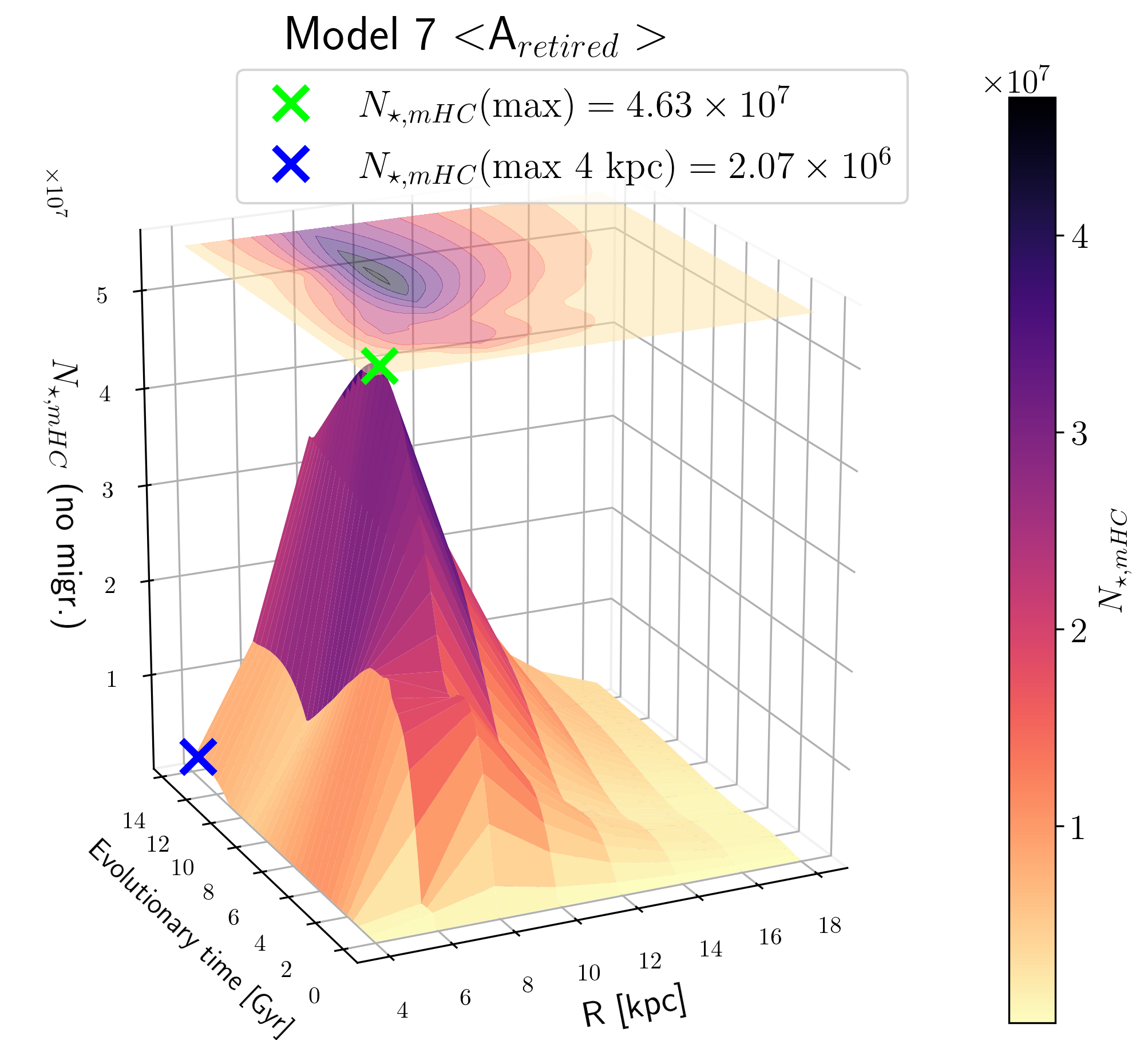}
     \caption{ Total number of retired  stars  having Earths ($N_{\star, mHC}$) as a function of the Galactocentric distance considering  different gas giant effects. In the upper panel   Model 1 (GG BAD) results  are reported and Model 7  (GG GOOD) in the lower one. Both models do not include the stellar migration.
  }
		\label{A_GOOD}
\end{figure}

\begin{figure}
	      \centering
\includegraphics[scale=0.52]{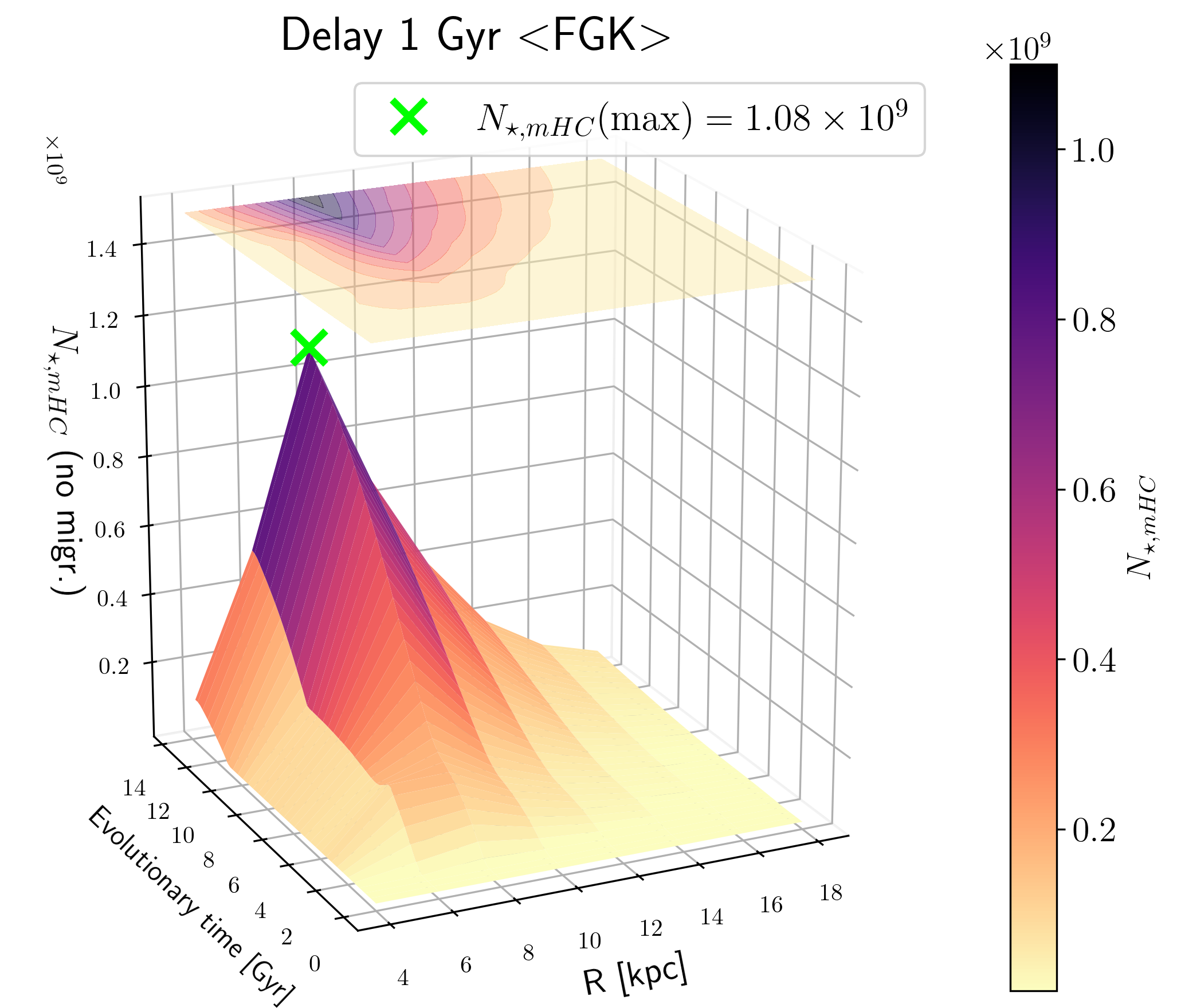}
     \includegraphics[scale=0.52]{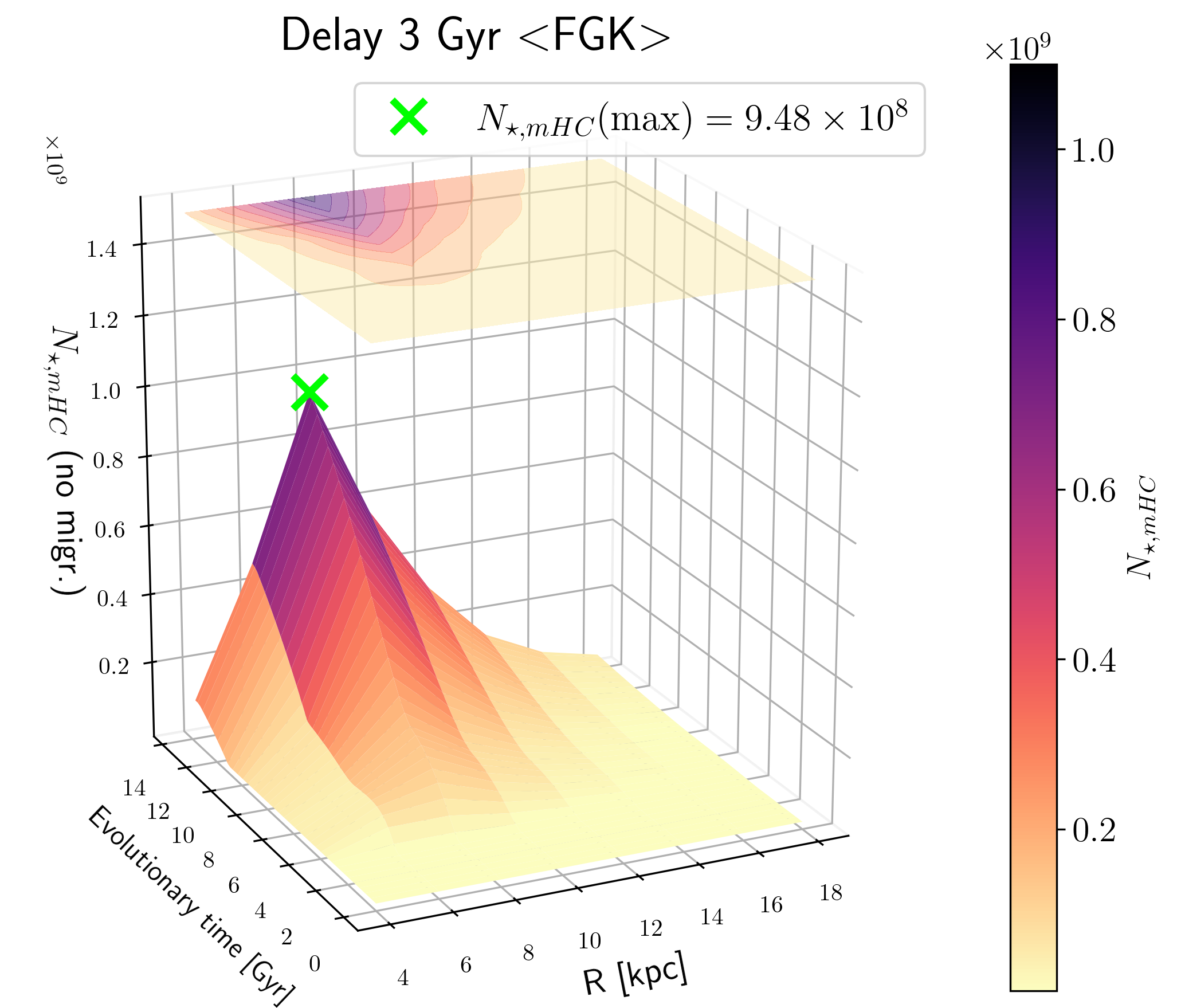}
     \caption{ As the upper panel of Fig. \ref{models1.2} (Model 1), but accounting for the time required to form planetary atmospheres. In the upper panel, it is assumed that 1 Gyr is needed for an atmosphere to form after the stellar host’s formation, while in the lower panel, this timescale is extended to 3 Gyr. } 
		\label{models_delay}
\end{figure}

In this Appendix we provide more results on the GHZ without stellar migration.
In Fig. \ref{A_GOOD},  we show the total number of retired A stars  having Earths ($N_{\star, mHC}$) as a function of the Galactocentric distance considering  different gas giant effects. In the upper panel   Model 1 (GG BAD) results  are reported and Model 7  (GG GOOD) in the lower one. In Fig. \ref{models_delay}, we show habitability maps  for a "modified Model 1", which accounts for the time necessary for a sustained increase in atmospheric O$_2$ to significant levels. On Earth, this process needed approximately 2.5 Gyr  \citep{Lyons2014}. 
Specifically, we evaluated delays of 1 Gyr and 3 Gyr.


\section{Observational data}  
\label{app:data}
\begin{figure}[t]
    \centering
\includegraphics[scale=0.17]{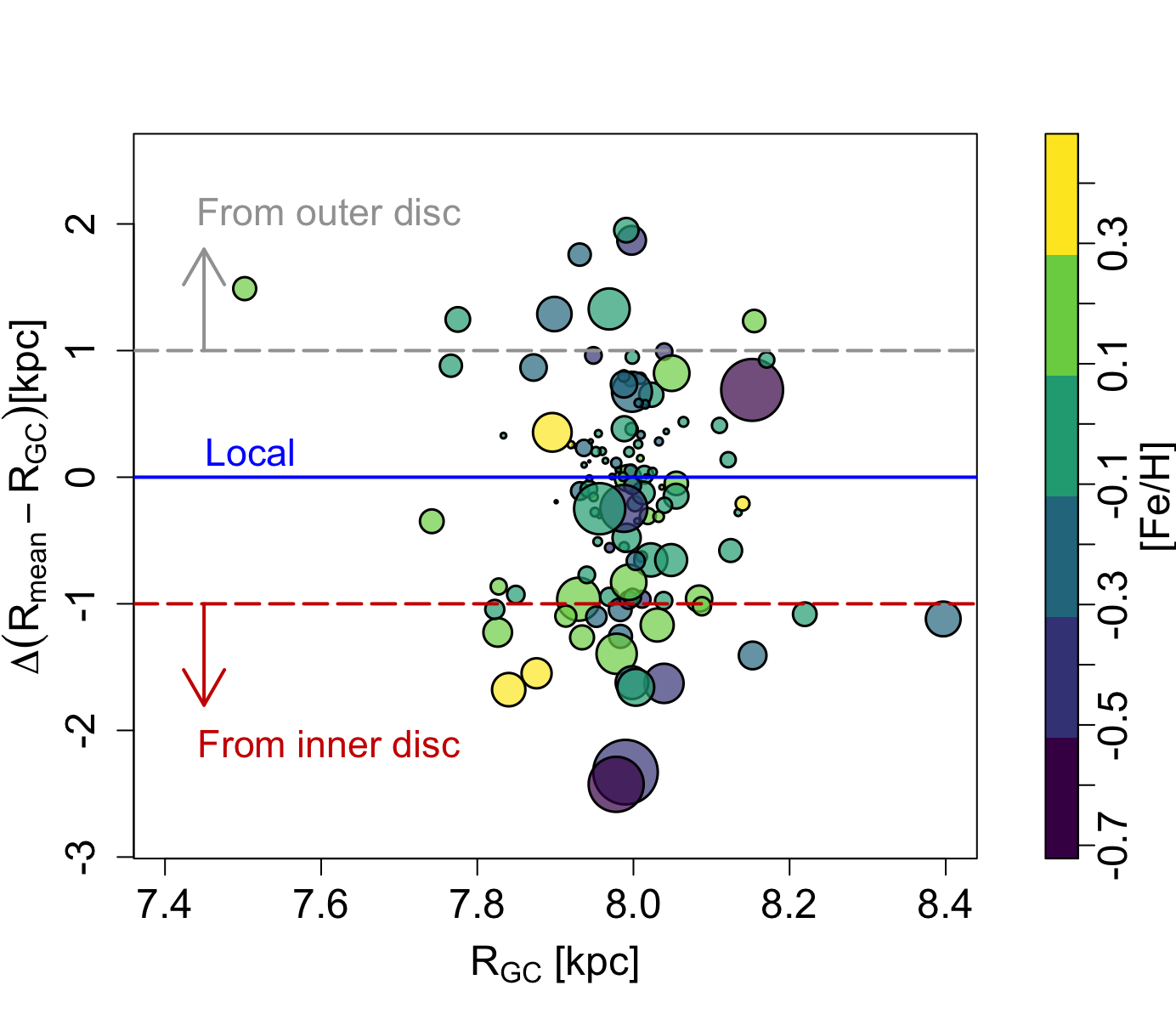}
    \caption{ Orbital radial variation ($\Delta(R_{mean}-R_{GC}$)) versus galactocentric distance ($R_{GC}$) of the sample of homogeneous determined stars. The marker size show the orbital eccentricity of the stars (i.e., smaller size means smaller eccentricity), while the color refers to the metallicity [Fe/H] as reported in the colourbar. The blue line marks a null difference between the R$_{mean}$
    of the orbit and the current R$_{GC}$, and marks the local region i.e., 8 $\pm$ 1 kpc. Grey line marks those stars that likely migrated from the outer disc ($\Delta >$ 1 kpc) while red dashed line marks those stars that likely migrated from the inner disc ($\Delta < -1$ kpc). }
    \label{fig:datamigration}
\end{figure}

In this Appendix, we provide detailed information about the the A$\&$S data sample used for the CMDF comparison with model predictions in  Fig. \ref{MDF_life_migr} and in Section \ref{res_migr}.
 From the Ariel Stellar Catalogue
\footnote{publicly available at \url{bit.ly/ArielStellarCatalogue}} \citep{MagriniAriel2022,t25}, we selected stars whose planet(s) have masses $M_P < 30\ M_\oplus$, for a total of 57 stars. Then, we complemented such a sample with the SWEET-Cat Catalogue
\footnote{publicly available at \url{https://sweetcat.iastro.pt}} \citep{sousa2021,Santos2013}, selecting stars that are not in the Ariel sample, for a total of 67 stars hosting low mass planets. Both of these catalogues provide precise and, most importantly, homogeneous stellar properties (effective temperature, surface gravity and [Fe/H]), obtained from high-signal to-noise and high resolution spectra, which are essential characteristics for conducting robust population studies (e.g., \citealt{adibekyan2019,danielski2022}). In particular, we note that the [Fe/H] determination within both catalogues has been tested to be consistent \citep{brucalassi2022}. Our final sample consists of a total of 124 stars: 106 FGK and 18 M type stars. \\
To set such stars within the Galactic context we computed the velocity components and the orbital parameters with the {\sc galpy} package \citep{Bovy2015}\footnote{\url{https://www.galpy.org/}}, using the {\em Gaia} DR3 data. We followed the same procedure as described in \cite{MagriniAriel2022} and \cite{t25}. Consequently, we compared their current Galactocentric position with the average position of their orbits (calculated as the average of apogalactic and perigalactic radii). We show in Fig. \ref{fig:datamigration} the orbital radial variation $(R_{mean}-R_{GC})$ versus their Galactocentric distance ($R_{GC}$). While the local population (i.e., those with variation of less than 1 kpc) is the only one showing very low eccentricities,  other more eccentric stars show larger variations, which indicates that they have probably originated in the outer or inner part of the Galactic disc. We used such radial variations to divide our stars within {\it local} and {\it inner} and {\it outer} populations.

\end{appendix}  

\end{document}